\documentclass[floatfix,twocolumn,showpacs,preprintnumbers,amsmath,amssymb,pra,superscriptaddress,longbibliography]{revtex4-1}
\usepackage{color}
\usepackage[usenames,dvipsnames,svgnames,table]{xcolor}
\usepackage[colorlinks=true,linkcolor=blue,urlcolor=blue,citecolor=blue]{hyperref}
\usepackage{mathtools}
\usepackage{graphicx}
\usepackage{dcolumn}
\usepackage{array}
\usepackage{lipsum}
\usepackage{bm}
\usepackage{subfigure}
\usepackage{amssymb}
\usepackage{multirow}
\usepackage{tabularx}
\usepackage{amsmath}
\usepackage{braket}
\graphicspath{{plots/}}
 \usepackage{lipsum}
\usepackage{mathrsfs}

\begin{document}
\title{
Overcoming finite-size effects in electronic structure simulations at extreme conditions
}

\author{Tobias Dornheim}
\email{t.dornheim@hzdr.de}

\affiliation{Center for Advanced Systems Understanding (CASUS), D-02826 G\"orlitz, Germany}

\author{Jan Vorberger}
\affiliation{Helmholtz-Zentrum Dresden-Rossendorf (HZDR), D-01328 Dresden, Germany}

\begin{abstract}
\textit{Ab initio} quantum Monte Carlo (QMC) methods in principle allow for the calculation of exact properties of correlated many-electron systems, but are in general limited to the simulation of a finite number of electrons $N$ in periodic boundary conditions. Therefore, an accurate theory of finite-size effects is indispensable to bridge the gap to realistic applications in the thermodynamic limit. In this work, we revisit the uniform electron gas (UEG) at finite temperature as it is relevant to contemporary research e.g. in the field of warm dense matter. In particular, we present a new scheme to eliminate finite-size effects both in the static structure factor $S(q)$ and in the interaction energy $v$, which is based on the density response formalism. We demonstrate that this method often allows to obtain $v$ in the TDL within a relative accuracy of $\sim0.2\%$ from as few as $N=4$ electrons without any empirical choices or knowledge of results for other values of $N$.
Finally, we evaluate the applicability of our method upon increasing the density parameter $r_s$ and decreasing the temperature $T$.
\end{abstract}

\maketitle

\section{Introduction\label{sec:introduction}}

The accurate numerical simulation of interacting many-electron systems constitutes a central challenge in many domains of theoretical physics, quantum chemistry, material science, and related disciplines~\cite{quantum_theory}. Yet, while the equations governing these systems have been known for approximately a century, there still does not exist a practical tool to solve them in many situations and one has to rely on approximations.

A particularly important method for this purpose is density functional theory (DFT), which has emerged as the de-facto work horse of quantum chemistry and boasts a remarkable success regarding the description of both bulk materials and molecules~\cite{Burke_Perspective_JCP_2012,Jones_RevModPhys_2015}. More specifically, DFT simulations are computationally efficient, as the the complicated $N$-body problem of interest is mapped on an effective single-body problem that can actually be solved numerically. Yet, this mapping comes at the expense of self-sufficiency, as information about electronic correlations must be supplied externally in the form of the in general unknown exchange--correlation functional.

In contrast, computationally more expensive quantum Monte Carlo (QMC) methods are, at least in principle, capable to provide an exact solution to the $N$-electron problem~\cite{Booth_Nature_2013,anderson2007quantum} without any a-priory information or empirical input. In particular, it was only the seminal QMC study of the uniform electron gas (UEG) by Ceperley and Alder~\cite{Ceperley_Alder_PRL_1980} which allowed for the subsequent construction of accurate electron exchange--correlation functionals~\cite{Perdew_Wang_PRB_1992,Perdew_Zunger_PRB_1981,vwn,PBE_1996} that made the success of DFT possible in the first place. At the same time, most QMC methods are by definition only applicable for a finite number of electrons $N$, while most physical systems are given in the thermodynamic limit (TDL), i.e., in the limit of $N\to\infty$ with the density $n=N/V$ remaining constant. This problem is substantially exacerbated by the notorious fermion sign problem~\cite{Loh_PRB_1990,troyer,dornheim_sign_problem}, which leads to an exponential increase of computation time with $N$, see Ref.~\cite{dornheim_sign_problem} for a topical review article. In fact, the sign problem has been revealed as $NP$-hard for a certain type of Hamiltonian by Troyer and Wiese~\cite{troyer}, and a general solution appears to be improbable.

For these reasons, the solid understanding of finite-size effects in QMC data and their subsequent estimation in the form of a finite-size correction (FSC) is of paramount importance for the QMC community and beyond, and constitutes a highly active topic of research~\cite{Fraser_Foulkes_PRB_1996,PhysRevB.59.1917,Williamson_PRB_1997,Lin_Zong_Ceperley_PRE_2001,Chiesa_PRL_2006,Chiesa_Proceeding_2007,Azadi_Foulkes_PRB_2019,Shepherd_JCP_2019,Holzmann_FSC_PRB_2016,Holzmann_2011,Drummond_PRB_2008,Holzmann_PRL_2011,dornheim_prl,review,moroni2,Krakauer_PRL_2008,2018EPJWC.18202044B,Spink_PRB_2013,Brown_PRL_2013,groth_jcp}.
Moreover, while most works have been devoted to the study of electrons at ambient conditions, i.e., in the ground state~\cite{Foulkes_RevModPhys_2001}, there has recently emerged a growing interest in the properties of matter at extreme densities and temperatures. Of particular importance is the regime of so-called warm dense matter~\cite{new_POP,wdm_book,review}, which is characterized by the simultaneous importance of thermal excitations, Coulomb correlations, and fermionic exchange-effects and naturally occurs e.g. in astrophysical objects such as giant planet interiors~\cite{Militzer_2008} and brown dwarfs~\cite{saumon1,becker}. In addition, warm dense matter has been predicted to occur on the pathway towards inertial confinement fusion~\cite{hu_ICF}, and is routinely realized experimentally in large research centres around the globe~\cite{falk_wdm}.

This interest has sparked a series of new developments in the field of electronic QMC simulations at finite temperature~\cite{Driver_Militzer_PRL_2012,Blunt_PRB_2014,dornheim_POP,Brown_PRL_2013,Dornheim_NJP_2015, Schoof_PRL_2015,Malone_JCP_2015,Militzer_Driver_PRL_2015,Malone_PRL_2016,dornheim_prl,dornheim_cpp,groth_jcp,dornheim_pre,Driver_PRE_2018,Dornheim_PRL_2020,Dornheim_JCP_2020,lee2020phaseless,Rubenstein_auxiliary_finite_T,Yilmaz_JCP_2020}, which, in turn, has caused the need to understand the impact of thermal excitations on finite-size effects. In general, this problem can be re-stated as the search for a short-range property that can be accurately inferred from a QMC simulation of a finite-system and, in combination with a readily available theory such as the random phase approximation (RPA) [see Eq.~(\ref{eq:chi}) below], yields the full description of a system in the TDL.

In the ground state, Chiesa \textit{et al.}~\cite{Chiesa_PRL_2006} have proposed to use the static structure factor (SSF) $S(q)$ for this purpose, which indeed has been shown empirically to only weakly depend on $N$ both at zero and finite temperature~\cite{dornheim_prl,review}. This finding has allowed to introduce a simple analytical first-order correction to the finite-size error of the interaction energy per particle $v$, which was subsequently generalized by Brown \textit{et al.}~\cite{Brown_PRL_2013} to arbitrary temperatures.
Yet, this correction breaks down both for high temperature and density and, thus, is inapplicable over substantial parts of the warm-dense matter regime. The full estimation of the main contribution to the finite-size error of $v$ then requires the estimation of a discretization error [see Eq.~(\ref{eq:FSC_v_d}) below] using a suitable trial function, like the SSF within RPA or a more sophisticated dielectric theory like the approximate scheme by Singwi \textit{et al.}~\cite{stls_original,stls,stls2} (STLS). This was demonstrated by Dornheim, Groth, and co-workers~\cite{dornheim_prl,dornheim_POP,review}, who showed that this procedure reduces the finite-size error in the QMC data by approximately two order of magnitude, allowing for a reliable subsequent extrapolation to the TDL~\cite{dornheim_prl,groth_prl}.

In the present work, we go one step further and address the source of the residual finite-size errors after the discretization error has been eliminated. To this end, we employ the density response formalism, and propose that in many cases, the static local field correction constitutes a more suitable choice for a short-range exchange--correlation function that can be accurately estimated from a QMC simulation of as few as $N=4$ particles. More specifically, our new scheme allows for a direct estimation of the finite-size error in $S(q)$ itself, which already constitutes an important finding in itself. Moreover, this FSC for the SSF can subsequently be used to eliminate the residual finite-size error in $v$, and we find that often $N=4$ particles are sufficient to estimate $v$ with a relative accuracy of $\sim0.2\%$ without any empirical input, extrapolation, or knowledge about QMC results for multiple $N$.

The paper is organized as follows: In Sec.~\ref{sec:theory}, we introduce the required theoretical background, starting with the density response formalism (\ref{sec:density_response}) and the fluctuation--dissipation theorem (\ref{sec:FDT}), followed by our new theory of finite-size effects in the SSF (\ref{sec:SA}), and the implications for the FSC of the interaction energy $v$ (\ref{sec:FSC_v}).
The application of our scheme is demonstrated in Sec.~\ref{sec:results}, starting with an extensive analysis at extreme density and temperature in Sec.~\ref{sec:high}.
In addition, we analyze the applicability of the method for the important cases of increasing coupling strength (\ref{sec:coupling}) and low temperature (\ref{sec:temperature}). The paper is concluded by a brief summary and outlook in Sec.~\ref{sec:outlook}.

\section{Theory\label{sec:theory}}

Throughout this work, we restrict ourselves to an unpolarized uniform electron gas (UEG), see Ref.~\cite{review} for details.
In addition, the UEG is characterized by two parameters: i) the density parameter $r_s=\overline{r}/a_\textnormal{B}$, where $\overline{r}$ and $a_\textnormal{B}$ denote the average particle distance and first Bohr radius, and ii) the degeneracy temperature $\theta=k_\textnormal{B}T/E_\textnormal{F}$, where $E_\textnormal{F}$ is the usual Fermi energy~\cite{Ott2018,quantum_theory}.

All formulas and results are given in Hartree atomic units.
\subsection{Density response and local field correction\label{sec:density_response}}

The density response of an electron gas to an external harmonic perturbation~\cite{Dornheim_PRL_2020} of wave-number $q$ and frequency $\omega$ is---within linear response theory---fully described by the dynamic density response function~\cite{quantum_theory,kugler1}
\begin{eqnarray}\label{eq:chi}
\chi(q,\omega) = \frac{\chi_0(q,\omega)}{1-\frac{4\pi}{q^2}\left[1-G(q,\omega)\right]\chi_0(q,\omega)}\ .
\end{eqnarray}
Here $\chi_0(q,\omega)$ denotes the density response function of the ideal Fermi gas and the information about exchange--correlation effects is contained in the dynamic local field correction $G(q,\omega)$.

Let us next consider the static limit, i.e.,
\begin{eqnarray}
\chi(q) = \lim_{\omega\to0}\chi(q,\omega) \ .
\end{eqnarray}
In this limit, accurate data for Eq.~(\ref{eq:chi}) have been presented by Dornheim \textit{et al.}~\cite{dornheim_ML,dornheim_electron_liquid,dornheim_HEDP} based on the relation~\cite{bowen2}
\begin{eqnarray}\label{eq:static_chi}
\chi({q}) = -n\int_0^\beta \textnormal{d}\tau\ F({q},\tau) \quad ,
\end{eqnarray}
with the imaginary-time density--density correlation function being defined as
\begin{eqnarray}\label{eq:F}
F(q,\tau) = \frac{1}{N} \braket{\rho(q,\tau)\rho(-q,0)}\ .
\end{eqnarray}
In addition, it is possible to obtain $\chi(q)$ from a simulation of the harmonically perturbed electron gas, which has been done both in the ground state~\cite{moroni,moroni2,bowen2} and at finite temperature~\cite{dornheim_pre,groth_jcp,Dornheim_PRL_2020}.
In principle, it is then straightforward to use $\chi(q)$ to solve Eq.~(\ref{eq:chi}) for the static LFC 
\begin{eqnarray}
G(q) &=& \lim_{\omega\to0}G(q,\omega) \nonumber\\ &=&
1 - \frac{q^2}{4\pi}\left( 
\frac{1}{\chi_0(q)} - \frac{1}{\chi(q)}
\right)\ .\label{eq:G_static}
\end{eqnarray}

For the present paper, it is important to explicitly indicate the dependence of different estimated properties on the system size, for which purpose we shall henceforth use the superscript $N$.
In particular, the most simple way to write Eq.~(\ref{eq:G_static}) is then given by
\begin{eqnarray}
G^N(q) =
1 - \frac{q^2}{4\pi}\left( 
\frac{1}{\chi^\textnormal{TDL}_0(q)} - \frac{1}{\chi^N(q)}
\right)\ .\label{eq:G_static_N}
\end{eqnarray}
Yet, we empirically know that $\chi^N(q)$ substantially depends on $N$, and this finite-size error is propagated into $G^N(q)$ as defined in Eq.~(\ref{eq:G_static_N}).
Fortunately, we also know that this error is almost exclusively due to the inconsistency between $\chi^\textnormal{TDL}_0(q)$ and $\chi^N(q)$, as the consistent determination of $G(q)$ would require us to instead use the density response function of the ideal system at the same system size, $\chi_0^N(q)$.
Thus, the finite-size corrected static local field correction can be computed as~\cite{moroni,moroni2}
\begin{eqnarray}\label{eq:G_static_FSC}
G^N_\textnormal{FSC}(q) &=&
1 - \frac{q^2}{4\pi}\left( 
\frac{1}{\chi^N_0(q)} - \frac{1}{\chi^N(q)}
\right)\\ \nonumber &\approx& G^\textnormal{TDL}(q)\ .
\end{eqnarray}
Moreover, we can readily define a FSC for $G(q)$,
\begin{eqnarray} \label{eq:FSC_for_G}
\Delta G^N(q) = \frac{q^2}{4\pi}\left(
\frac{1}{\chi_0^\textnormal{TDL}(q)} - \frac{1}{\chi_0^N(q)}
\right)\ ,
\end{eqnarray}
such that
\begin{eqnarray}\label{eq:FSC_G}
G^N_\textnormal{FSC}(q) = G^N(q) + \Delta G^N(q)\ .
\end{eqnarray}

\subsection{Fluctuation--dissipation theorem\label{sec:FDT}}

The fluctuation--dissipation theorem~\cite{quantum_theory}
\begin{eqnarray}\label{eq:FDT}
S({q},\omega) = - \frac{ \textnormal{Im}\chi({q},\omega)  }{ \pi n (1-e^{-\beta\omega})}
\end{eqnarray}
relates Eq.~(\ref{eq:chi}) to the dynamic structure factor $S(q,\omega)$ and, thus, directly connects the LFC to different material properties. 
In particular, the static structure factor is defined as the normalization of the DSF
\begin{eqnarray}\label{eq:Sq}
S(q) = \int_{-\infty}^\infty \textnormal{d}\omega\ S(q,\omega)\ ,
\end{eqnarray}
and thus entails an averaging over the full frequency range.
We stress that this is in contrast to the static density response function $\chi(q)$ introduced in the previous section, which is defined as the limit of $\omega\to0$.
The SSF, in turn, gives direct access to the interaction energy of the system, and for a uniform system it holds~\cite{review}
\begin{eqnarray}\label{eq:v}
v = \frac{1}{\pi} \int_0^\infty \textnormal{d}q\ \left[
S(q)-1
\right]\ .
\end{eqnarray}
Finally, we mention the adiabatic connection formula~\cite{review,groth_prl,ksdt}
\begin{eqnarray}\label{eq:adiabatic}
f_{xc}(r_s,\theta) = \frac{1}{r_s^2} \int_0^{r_s} \textnormal{d}\overline{r}_s\ v(\overline{r}_s,\theta)\overline{r}_s\ ,
\end{eqnarray}
which implies that the free energy $f$ (and, equivalently the partition function $Z$) can be inferred from either $\chi(q,\omega)$, $S(q,\omega)$, or $S(q)$.

\subsection{Finite-size correction of $S(q)$\label{sec:SA}}

Since the full frequency-dependence of $G(q,\omega)$ remains unknown in most cases, one might neglect dynamic effects and simply substitute $G(q)$ in Eq.~(\ref{eq:chi}). This leads to the dynamic density response function within the \emph{static approximation}~\cite{dornheim_dynamic,Hamann_PRB_2020,Dornheim_PRE_2020,dynamic_folgepaper},
\begin{eqnarray}\label{eq:static_approximation}
\chi_\textnormal{stat}(q,\omega) = \frac{\chi_0(q,\omega)}{1-\frac{4\pi}{q^2}\left[1-G(q)\right]\chi_0(q,\omega)}\ ,
\end{eqnarray}
which entails the frequency-dependence on an RPA level, but exchange-correlation effects are incorporated statically. Indeed, it has been shown that Eq.~(\ref{eq:static_approximation}) allows for an accurate (though not exact) description of many material properties~\cite{dornheim_dynamic,dynamic_folgepaper,dornheim_PRL_ESA_2020,Hamann_PRB_2020}. This finding constitutes the motivation for the present scheme to overcome finite-size effects in $S(q)$ and related quantities.

Very recently, Dornheim and Moldabekov~\cite{Dornheim_PRB_ESA_2021} have introduced the concept of an effectively frequency-averaged LFC $\overline{G}(q)$, i.e., a static LFC that is not defined in the limit of $\omega\to0$, but effectively incorporates the dependence on $\omega$.
For example, we can define such a LFC in a way that, when it is being inserted into Eqs.~(\ref{eq:static_approximation}), (\ref{eq:FDT}) and (\ref{eq:Sq}), exactly reproduces the QMC data for $S^N(q)$,
\begin{eqnarray}\label{eq:invert}
\overline{G}^N_\textnormal{invert}(q) = \textnormal{min}_{\overline{G}}\left(
\left|
S_{\overline{G}}(q) - S^N(q)
\right|
\right)\ .
\end{eqnarray}
In practice, we determine $\overline{G}^N_\textnormal{invert}(q)$
numerically by scanning the corresponding results for $S(q)$ over a dense grid of $\overline{G}$ for each particular wave number $q$.
In fact, it has been shown~\cite{dornheim_PRL_ESA_2020} that $G^N(q)$ is almost exactly equal to $\overline{G}^N_\textnormal{invert}(q)$ for wave numbers $q\lesssim3q_\textnormal{F}$. It is therefore reasonable to approximate the (a-priori unknown) FSC for $\overline{G}^N_\textnormal{invert}(q)$ by the FSC for the static limit $\Delta G^N(q)$ defined in Eq.~(\ref{eq:FSC_G}) above,
\begin{eqnarray}
\overline{G}^\textnormal{TDL}_\textnormal{invert}(q) \approx \overline{G}^\textnormal{FSC}_\textnormal{invert}(q) = \overline{G}^N_\textnormal{invert}(q) + \Delta G^N(q) \ . \label{eq:FSC_G_invert}
\end{eqnarray}
The resulting corrected value $\overline{G}^\textnormal{FSC}_\textnormal{invert}(q)$ is then used to compute the finite-size corrected value of the static structure factor $S_N^\textnormal{FSC}(q)$, again via Eqs.~(\ref{eq:static_approximation}), (\ref{eq:FDT}) and (\ref{eq:Sq}). The corresponding value of the FSC for $S^N(q)$ is then given by
\begin{eqnarray}\label{eq:FSC_S}
\Delta S^N(q) = S_N^\textnormal{FSC}(q) - S^N(q)\ .
\end{eqnarray}

We stress that our approach is inherently superior to the \emph{static approximation} introduced in Refs.~\cite{dornheim_PRL_ESA_2020,dornheim_dynamic,Hamann_PRB_2020}, as we only use it to estimate the finite-size error in $\overline{G}_\textnormal{invert}^N(q)$, which is comparably small. Thus, the full frequency-dependence of $G(q,\omega)$ is consistently incorporated into the definition of $\overline{G}_\textnormal{invert}^N(q)$, whereas we only neglect it for the FSC in Eq.~(\ref{eq:FSC_G_invert}).

\subsection{Finite-size correction of the interaction energy\label{sec:FSC_v}}

The interaction energy in the TDL is defined by evaluating the integral in Eq.~(\ref{eq:v}) above using the SSF in the TDL, $S^\textnormal{TDL}(q)$,
\begin{eqnarray}\label{eq:v_TDL}
v = \frac{1}{\pi} \int_0^\infty \textnormal{d}q\ \left[
S^\textnormal{TDL}(q)-1
\right]\ .
\end{eqnarray}

In contrast, the corresponding energy per particle for a finite system is defined as a sum over the reciprocal vectors $\mathbf{G}$ of the simulation cell,
\begin{eqnarray}\label{eq:v_N}
\frac{V^N}{N} = \frac{1}{2V}\sum_{\mathbf{G}\neq\mathbf{0}}\left[S^N(\mathbf{G})-1 \right] \frac{4\pi}{G^2} + \frac{\xi_\textnormal{M}}{2}\ ,
\end{eqnarray}
where $\xi_\textnormal{M}$ is the usual Madelung constant taking into account the interaction of a charge with its own background and periodic array of images~\cite{Fraser_Foulkes_PRB_1996}.

Following a short analysis that has been presented elsewhere~\cite{Drummond_PRB_2008,Chiesa_PRL_2006,dornheim_prl,review}, it can be seen that there are two potential sources for the difference between Eqs.~(\ref{eq:v_TDL}) and (\ref{eq:v_N}):
(i) the discretization error $\Delta v^N_\textnormal{d}$ due to the approximation of a continuous integral by a discrete sum; (ii) the error intrinsic in $S^N(q)\neq S^\textnormal{TDL}(q)$, leading to $\Delta v^N_\textnormal{i}$.

It is well known that (i) is the dominant effect (at least for a sufficiently large system-size $N$), and it can be accurately estimated using a decent trial function for $S(q)$, e.g. using STLS~\cite{stls,stls2,dornheim_prl}, or even RPA,
\begin{eqnarray}\label{eq:FSC_v_d}
\Delta v_\textnormal{d}^N\left[S_\textnormal{trial}(q)\right]  &=& \frac{1}{\pi} \int_0^\infty \textnormal{d}q\ \left[
S_\textnormal{trial}(q)-1
\right] \\ \nonumber & & - \left(
\frac{1}{2V}\sum_{\mathbf{G}\neq\mathbf{0}}\left[S_\textnormal{trial}(\mathbf{G})-1 \right] \frac{4\pi}{G^2} + \frac{\xi_\textnormal{M}}{2}
\right)\ .
\end{eqnarray}
In addition, we mention that the first-order contribution to Eq.~(\ref{eq:FSC_v_d}) is a direct consequence of the missing $\mathbf{G}=\mathbf{0}$ term in Eq.~(\ref{eq:v_N}), and can be exactly evaluated in RPA, which gives~\cite{Chiesa_PRL_2006,Brown_PRL_2013,review}
\begin{eqnarray}\label{eq:BCDC}
\Delta v^N_{\textnormal{d,0}} = \frac{\omega_p}{4N}\textnormal{coth}\left( \frac{\beta\omega_p}{2} \right)\ ,
\end{eqnarray}
where $\omega_p=\sqrt{r_s^3/3}$ denotes the usual plasma frequency.

Yet, often (ii), too, is not negligible, which is particularly true at either high temperature/small $r_s$ or in the low-temperature regime where $S^N(q)$ is afflicted with momentum shell effects~\cite{Spink_PRB_2013}. Moreover, the fermion sign problem~\cite{dornheim_sign_problem,troyer} still prevents QMC simulations except for very small $N$ for substantial parts of the WDM regime, which makes the accurate estimation of the intrinsic error highly desirable.

We thus make use of the new FSC for $S(q)$ given in Eq.~(\ref{eq:FSC_S}) above, and write the intrinsic finite-size error of $v$ as
\begin{eqnarray}\label{eq:FSC_v_i}
\Delta v_\textnormal{i}^N = \frac{1}{2V}\sum_{\mathbf{G}\neq\mathbf{0}}\left[S_N^\textnormal{FSC}(\mathbf{G}) - S^N(\mathbf{G}) \right] \frac{4\pi}{G^2}\ .
\end{eqnarray}

The fully finite-size corrected estimate for the interaction energy is then given by
\begin{eqnarray}\label{eq:FSC_v}
v^\textnormal{FSC} = \frac{V^N}{N} + \Delta v^N_\textnormal{d}\left[ S_\textnormal{trial}(q) \right]
+ \Delta v^N_\textnormal{i}\ .
\end{eqnarray}

\section{Results\label{sec:results}}

All results presented in this work have been obtained for the unpolarized UEG, i.e., with the number of spin-up and spin-down electrons being equal.

\subsection{High densities and temperatures\label{sec:high}}

\begin{figure}\centering\hspace*{-0.017\textwidth}
\includegraphics[width=0.48\textwidth]{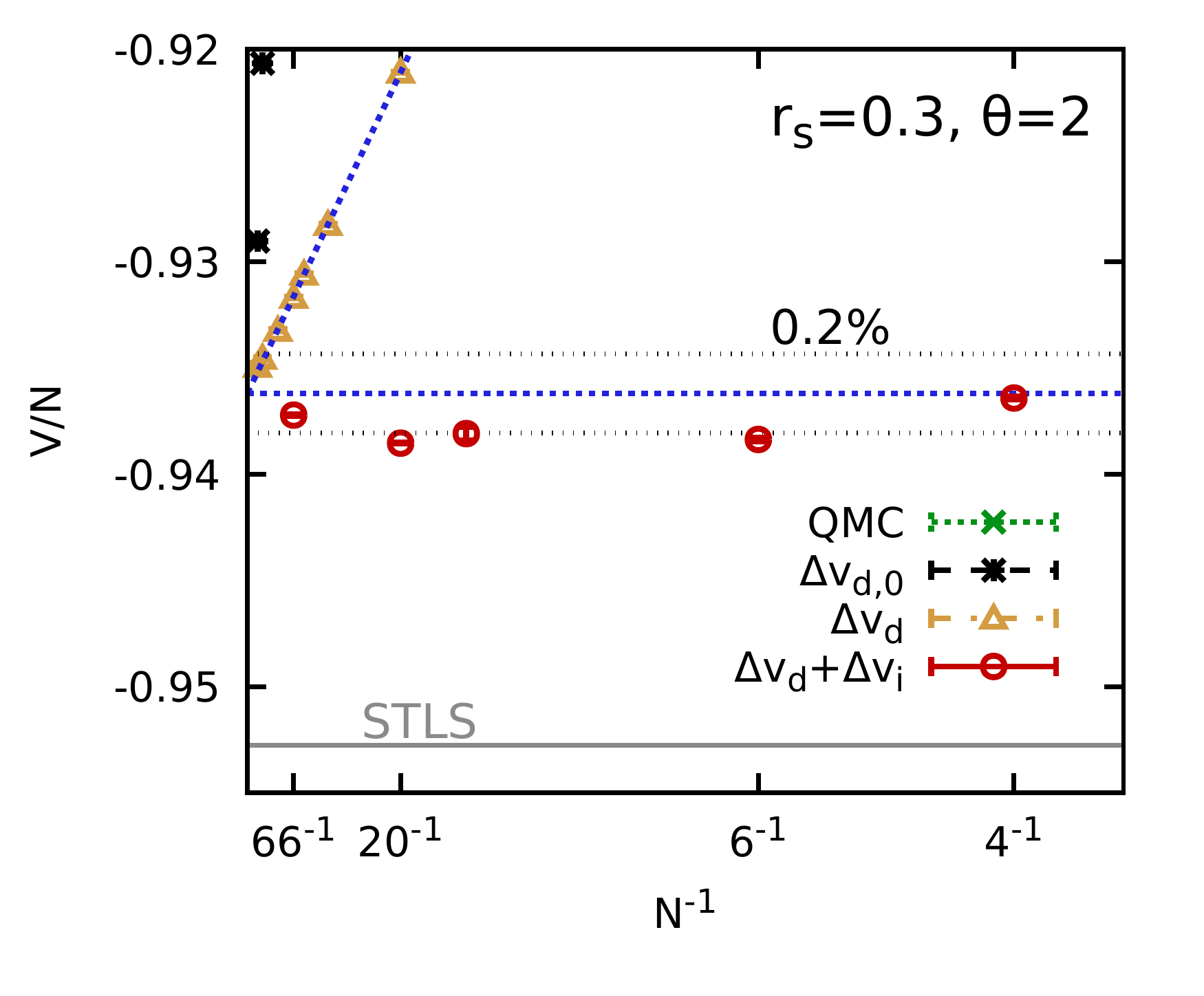}\\ \vspace*{-1.33cm}
\includegraphics[width=0.465\textwidth]{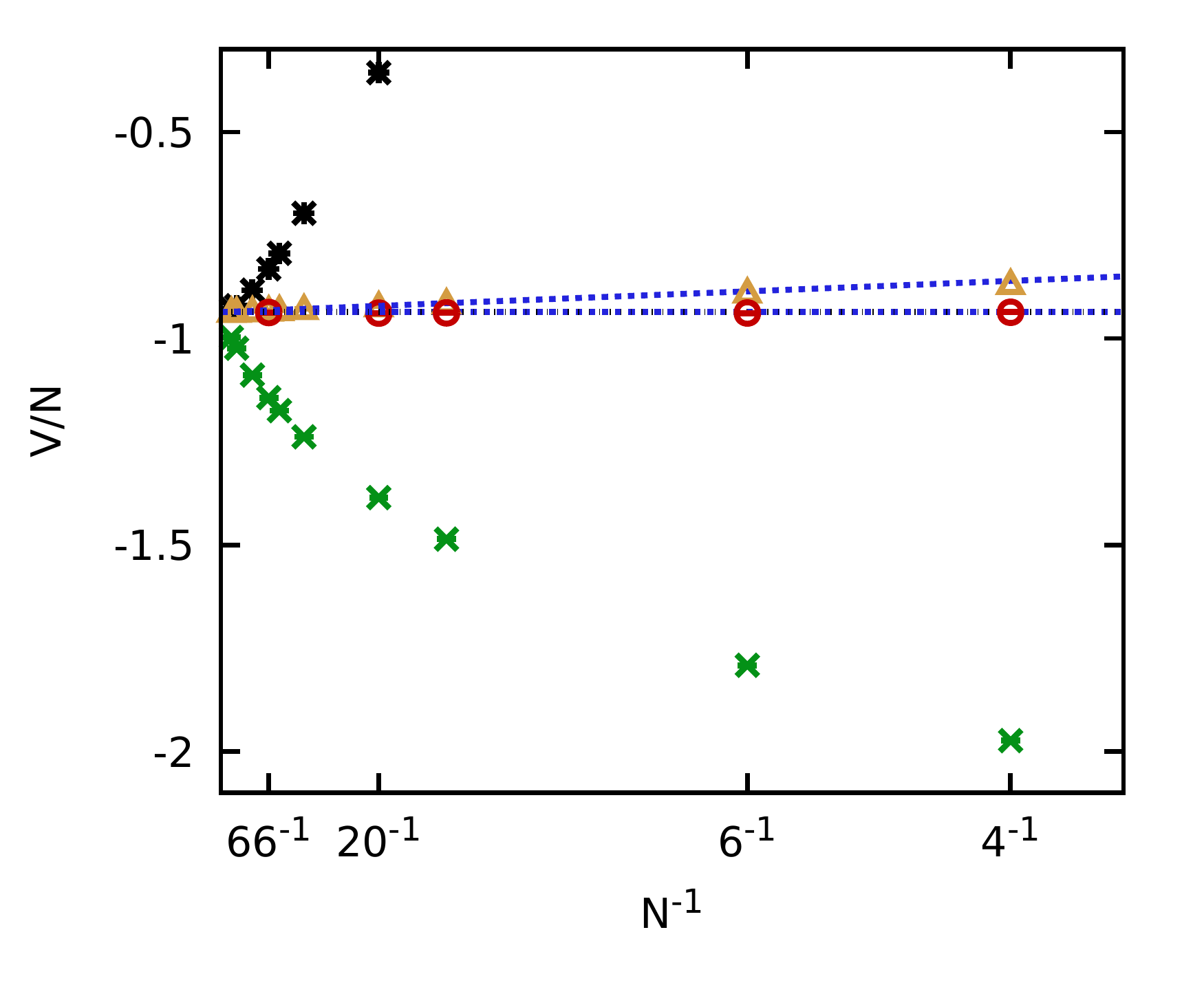}
\caption{\label{fig:v_rs0p3_theta2}
Interaction energy $v$ for $r_s=0.3$ and $\theta=2$. Top and bottom: zoomed and full view of $N$-dependence. Green crosses: raw CPIMC~\cite{dornheim_prl} and PIMC data; grey stars: first-order correction $\Delta v^N_{\textnormal{d},0}$, Eq.~(\ref{eq:BCDC}); yellow triangles: $\Delta v^N_\textnormal{d}$, Eq.~(\ref{eq:FSC_v_d}); red:  $\Delta v^N_\textnormal{d}+\Delta v^N_\textnormal{i}$, Eq.~(\ref{eq:FSC_v_i}); dotted blue: linear extrapolation; solid grey: STLS~\cite{stls_original,stls,stls2}.
}
\end{figure}

Let us start our investigation with an example at high density ($r_s=0.3$) and extreme temperature ($\theta=2$), which is close to the conditions in the solar core~\cite{fortov_review}.
In Fig.~\ref{fig:v_rs0p3_theta2}, we show the corresponding dependence of the interaction energy per particle $V^N/N$ on the system size $N$. Let us first consider the bottom panel, where we show the entire relevant energy-range. In particular, the green crosses depict the raw, uncorrected QMC data [partly Configuration path integral Monte Carlo (CPIMC)~\cite{groth_prb_2016,dornheim_prb_2016} data taken from Ref.~\cite{dornheim_prl}, and a few data points obtained with standard path integral Monte Carlo (PIMC)~\cite{boninsegni1,boninsegni2,cep}] that have been computed from Eq.~(\ref{eq:v_N}). Evidently, these data exhibit a strong dependence on $N$ reaching approximately $100\%$ for $N=4$. Even for the largest depicted system size, $N=300$, finite size errors have not completely vanished.
The yellow triangles have been obtained by adding to the raw data the FSC for the discretization error using $S^\textnormal{STLS}(q)$ as a trial function; see Eq.~(\ref{eq:FSC_v_d}) above. This leads to a remarkable reduction of the finite-size errors in the data set, and the residual error appears to be of the order of $\sim1\%$ even for as few as $N=4$ particles. This validates the previous conclusion from Ref.~\cite{dornheim_prl} that $\Delta v^N_\textnormal{d}$ does indeed constitute the dominant contribution to the total finite-size error $\Delta v^N$.
For completeness, we also show the first-order correction $\Delta v^N_{\textnormal{d},0}$ [Eq.~(\ref{eq:BCDC})], which is depicted by the dark grey stars. Yet, this first-order correction is only reasonable for very large numbers of electrons $N$, and even leads to an increase in finite-size errors for $N\lesssim20$.
Finally, the dotted blue line depicts a linear fit to the yellow triangles for $N\geq20$, which has been used to extrapolate the residual error of these data in Refs.~\cite{dornheim_POP,dornheim_prl,review,groth_prl}.

\begin{figure}\centering\hspace*{0.01\textwidth}
\includegraphics[width=0.462\textwidth]{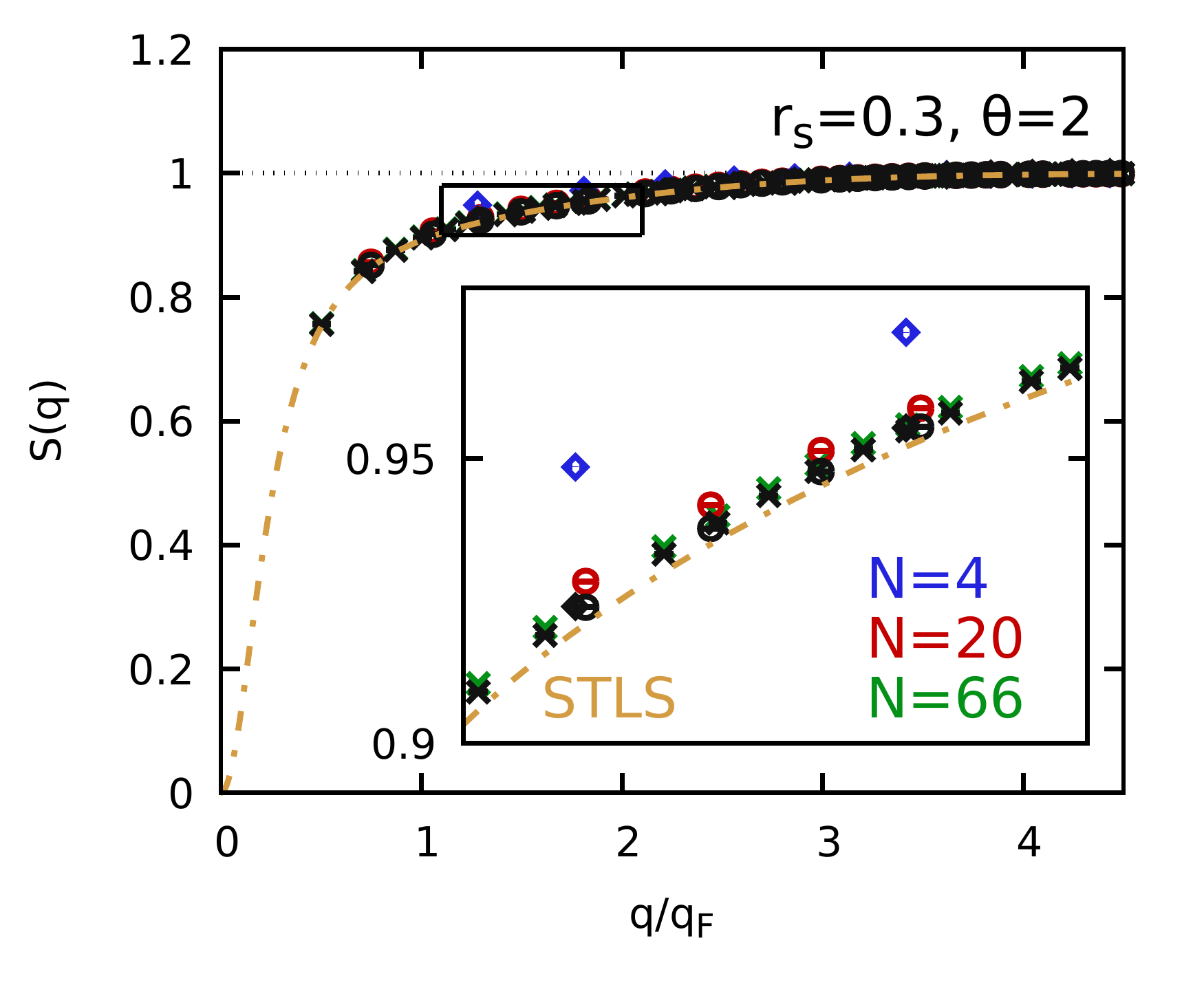}
\caption{\label{fig:Sq_rs0p3_theta2}
Static structure factor for $r_s=0.3$ and $\theta=2$. Coloured symbols depict raw QMC data for $N=66$ (green crosses), $N=20$ (red circles), and $N=4$ (blue diamonds). The corresponding black symbols have been finite-size corrected using Eq.~(\ref{eq:FSC_S}). Dash-dotted yellow: STLS~\cite{stls_original,stls,stls2}. Note that $q_\textnormal{F}$ denotes the Fermi wave number~\cite{quantum_theory}.
}
\end{figure}

Let us postpone the discussion of the red circles in Fig.~\ref{fig:v_rs0p3_theta2}, and instead focus on the source of the remaining $N$-dependence after the discretization error has been eliminated. To this end, we show QMC results for the static structure factor in Fig.~\ref{fig:Sq_rs0p3_theta2} at the same conditions for three values of $N$. More specifically, the coloured symbols depict the raw, uncorrected QMC results for $S^N(q)$ for $N=4$ (blue diamonds), $N=20$ (red circles), and $N=66$ (green crosses). First and foremost, we note that the dependence of $S^N(q)$ on $N$ is indeed small when compared to the substantially more pronounced finite-size error of $V^N/N$ itself. Still, there do appear significant differences between the data for different $N$, which is particularly pronounced for $N=4$, as it is expected.

Adding the FSC $\Delta S^N(q)$ from Eq.~(\ref{eq:FSC_S}) to $S^N(q)$ leads to the black symbols in Fig.~\ref{fig:Sq_rs0p3_theta2}. Evidently, this correction works well at these conditions, and the black diamonds, squares, and crosses can hardly be distinguished with the naked eye. This, in turn, means that as few as $N=4$ particles are sufficient to obtain a wave-number resolved description of electronic correlations in the UEG, despite the apparently large finite-size effects at such an extreme density.

Having thus verified our theory for finite-size errors in the static structure factor, we can readily compute the second source of finite-size errors in the interaction energy, namely the intrinsic error $\Delta v^N_\textnormal{i}$, Eq.~(\ref{eq:FSC_v_i}). The results of removing both intrinsic and discretization errors are depicted by the red circles in Fig.~\ref{fig:v_rs0p3_theta2}. In particular, our new procedure leads to a striking reduction in finite-size errors even for $N=4$ electrons, without any empirical input or knowledge about the behaviour for other $N$. Further, the corrected data for all $N$ are within $\sim0.2\%$ of the value that was obtained from the extrapolation of the yellow triangles.
For comparison, we have also included the prediction for $v$ from STLS~\cite{stls,stls2} (horizontal grey line) that is obtained by inserting $S^\textnormal{STLS}(q)$ in Eq.~(\ref{eq:v_TDL}).
Even though STLS and related dielectric theories are often considered to be rather accurate at these conditions~\cite{dornheim_HEDP}, we can clearly resolve a systematic overestimation of $v$ by $\sim1\%$. Yet, this deficiency has no impact on the estimation of $\Delta^N_\textnormal{d}[S^\textnormal{STLS}(q)]$, since the systematic error in $S^\textnormal{STLS}(q)$ cancels in Eq.~(\ref{eq:FSC_v_d}).

\begin{figure}\centering\hspace*{0.01\textwidth}
\includegraphics[width=0.462\textwidth]{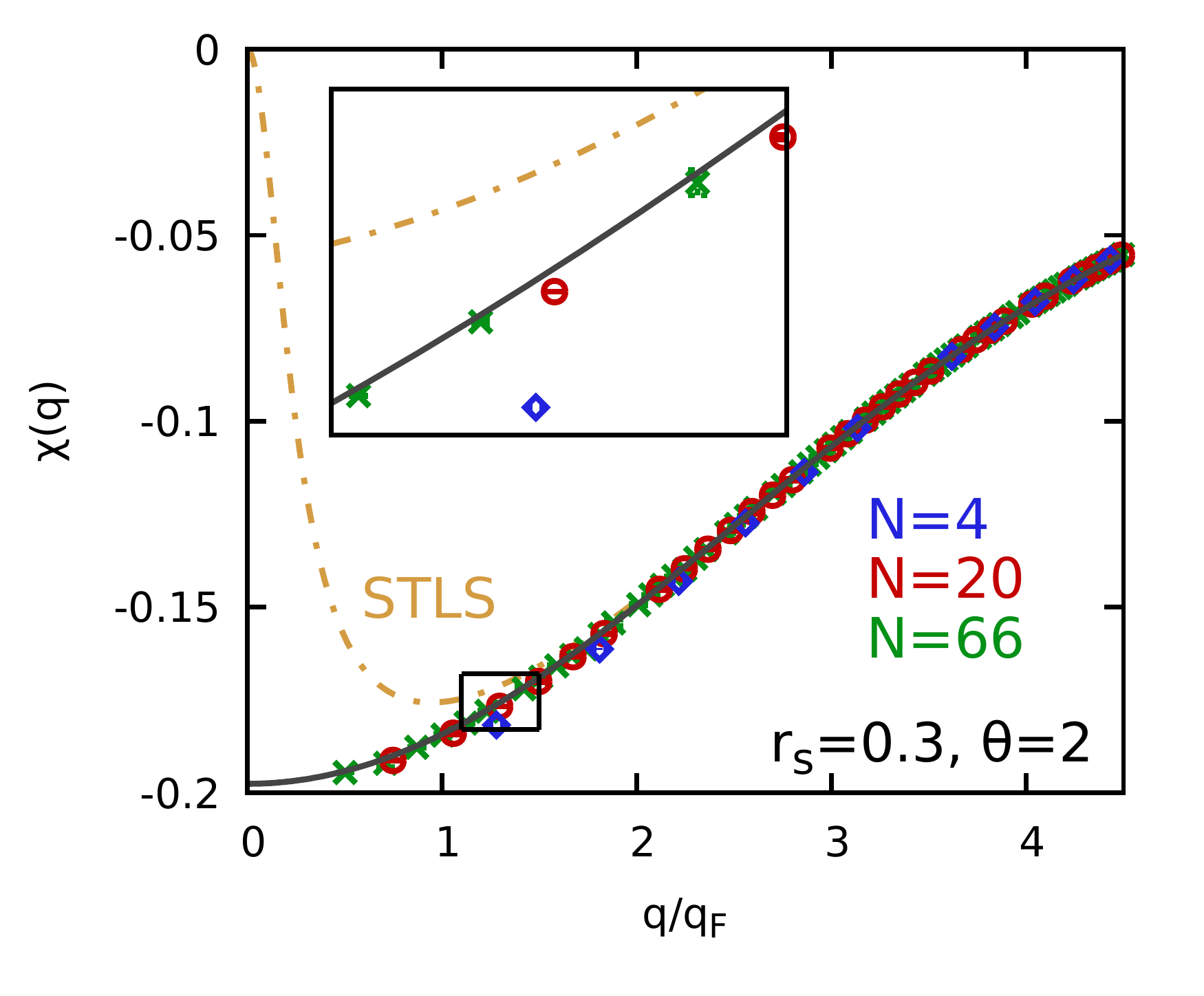}
\caption{\label{fig:Chi_rs0p3_theta2}
Static density response function $\chi(q)$ for $r_s=0.3$ and $\theta=2$. Coloured symbols depict CPIMC data~\cite{groth_jcp} for $\chi^N_0(q)$ for $N=66$ (green crosses), $N=20$ (red circles), and $N=4$ (blue diamonds). Solid black: $\chi^\textnormal{TDL}_0(q)$; Dash-dotted yellow: STLS~\cite{stls_original,stls,stls2}.
}
\end{figure}

To further illustrate the idea behind our new FSC-scheme, we show the density response function of the corresponding noninteracting (ideal) Fermi system in Fig.~\ref{fig:Chi_rs0p3_theta2}. In particular, the solid black curve shows $\chi^\textnormal{TDL}_0(q)$, and the blue diamonds, red circles, and green crosses depict $\chi^N_0(q)$ for $N=4$, $N=20$, and $N=66$, respectively. Evidently, there appear significant deviations between $\chi^\textnormal{TDL}_0$ and $\chi_0^N$ that closely resemble the trend observed in $S^N(q)$ shown in Fig.~\ref{fig:Sq_rs0p3_theta2} above. In addition, we have also included $\chi^\textnormal{STLS}(q)$ in the figure (dash-dotted yellow curve), which serves as a guide to the eye.

\begin{figure}\centering
\includegraphics[width=0.465\textwidth]{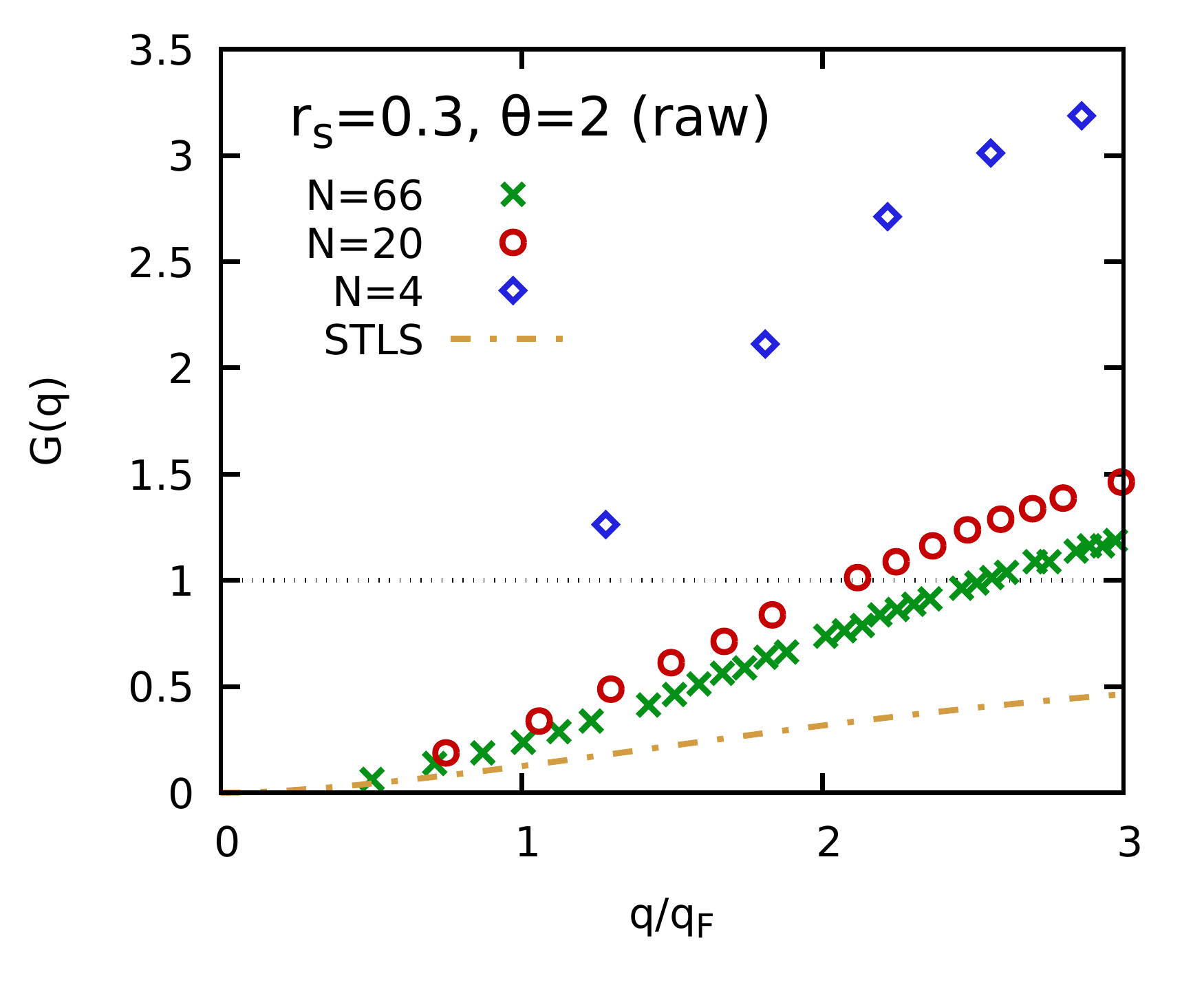}\\ \vspace*{-1.33cm}
\includegraphics[width=0.465\textwidth]{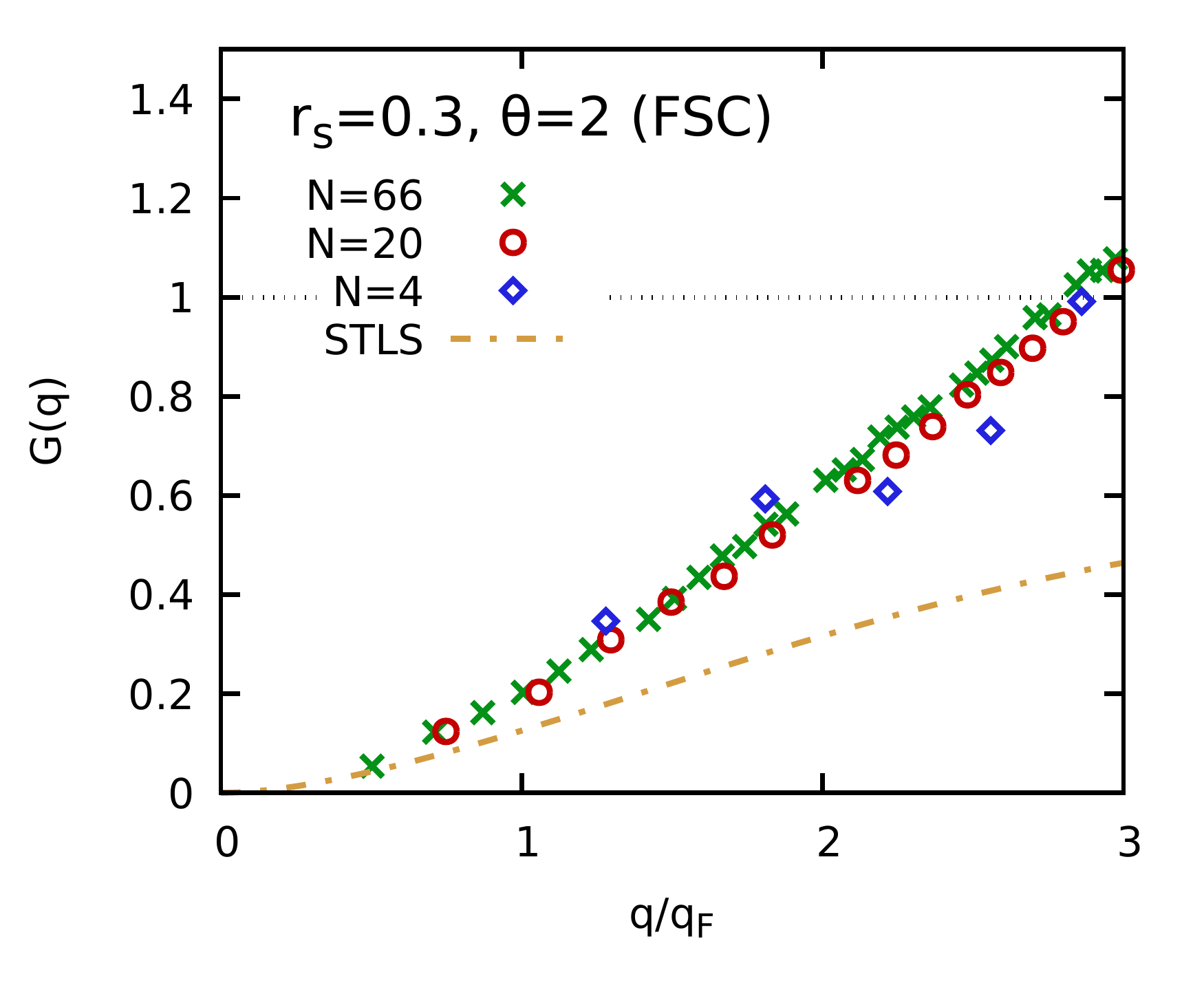}
\caption{\label{fig:G_rs0p3_theta2}
Effective local field correction $\overline{G}^N_\textnormal{invert}(q)$ for $r_s=0.3$ and $\theta=2$, top: raw (uncorrected) results, see Eq.~(\ref{eq:invert}); bottom: FSC-corrected data, see Eq.~(\ref{eq:FSC_G_invert}).
}
\end{figure}

To conclude our analysis for the present example, we show the actual effectively static local field correction $\overline{G}^N_\textnormal{invert}(q)$ in Fig.~\ref{fig:G_rs0p3_theta2} for the same particle numbers. The top panel shows the uncorrected raw data, and we see substantial finite-size effects for all data sets exceeding $100\%$ for $N=4$.
The bottom panel shows the corresponding results for $\overline{G}^\textnormal{FSC}(q)$ [see Eq.~(\ref{eq:FSC_G_invert})], and we find that the correction $\Delta G^N(q)$ defined in Eq.~(\ref{eq:FSC_for_G}) removes nearly the entire dependence on $N$.
This finding can be interpreted in the following way: the local field corrections $G(q,0)$ and $\overline{G}_\textnormal{invert}(q)$ have been defined as the difference between the exact density response / static structure factor of the system, and the prediction of mean-field theory, i.e., RPA. Therefore, it is a short-range function taking into account exchange--correlation effects, which are hardly affected by the system size at these conditions. Correspondingly, we can accurately extract either $G$ or $\overline{G}_\textnormal{invert}$ from a single simulation of the UEG even for $N=4$. Finally, this information about the short-range behaviour can be combined with the mean-field description, thus yielding the full, wave-number resolved description of the system in the TDL.

\begin{figure}\centering\hspace*{-0.033\textwidth}\includegraphics[width=0.505\textwidth]{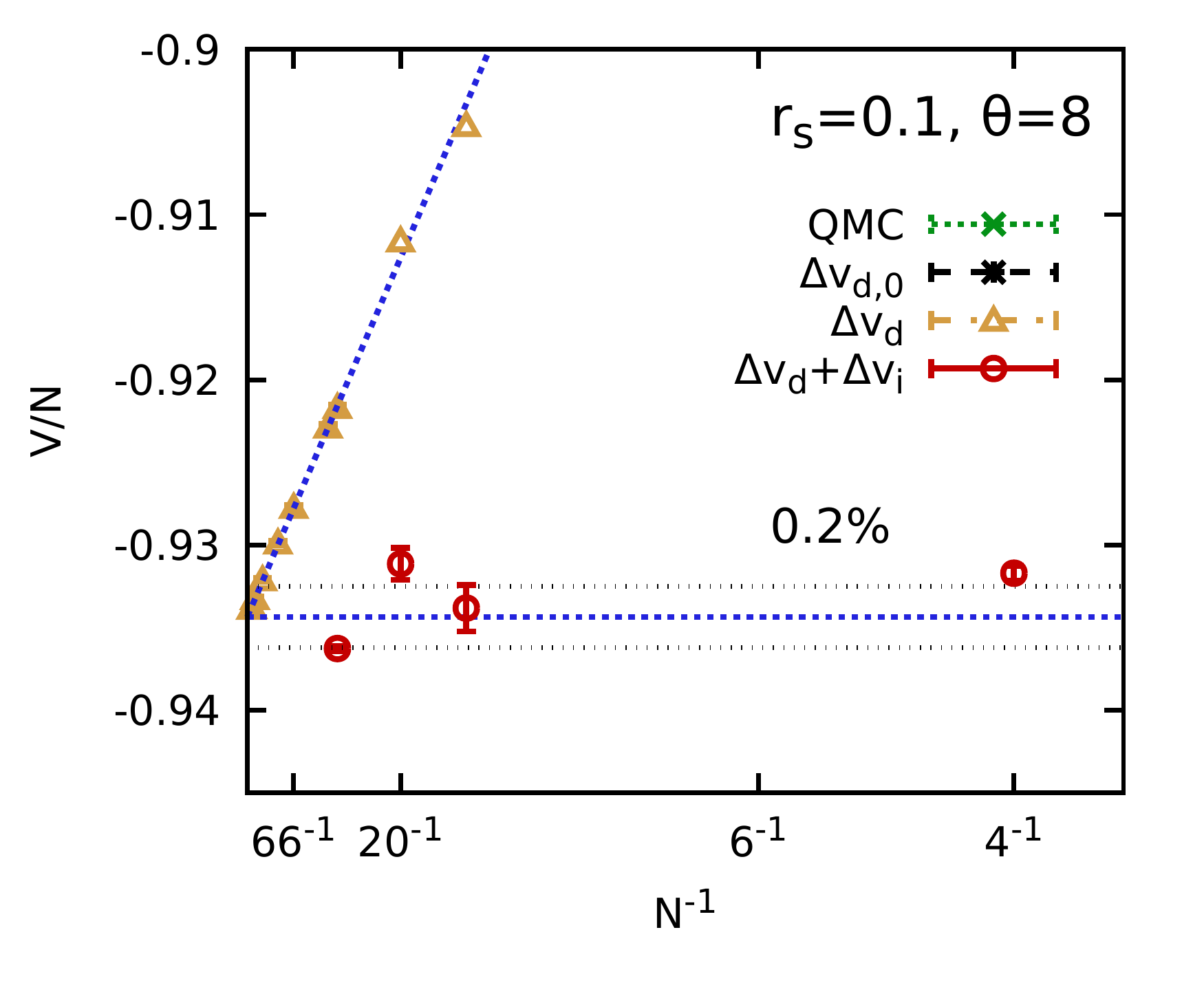}\\\vspace{-1.4cm}
\includegraphics[width=0.465\textwidth]{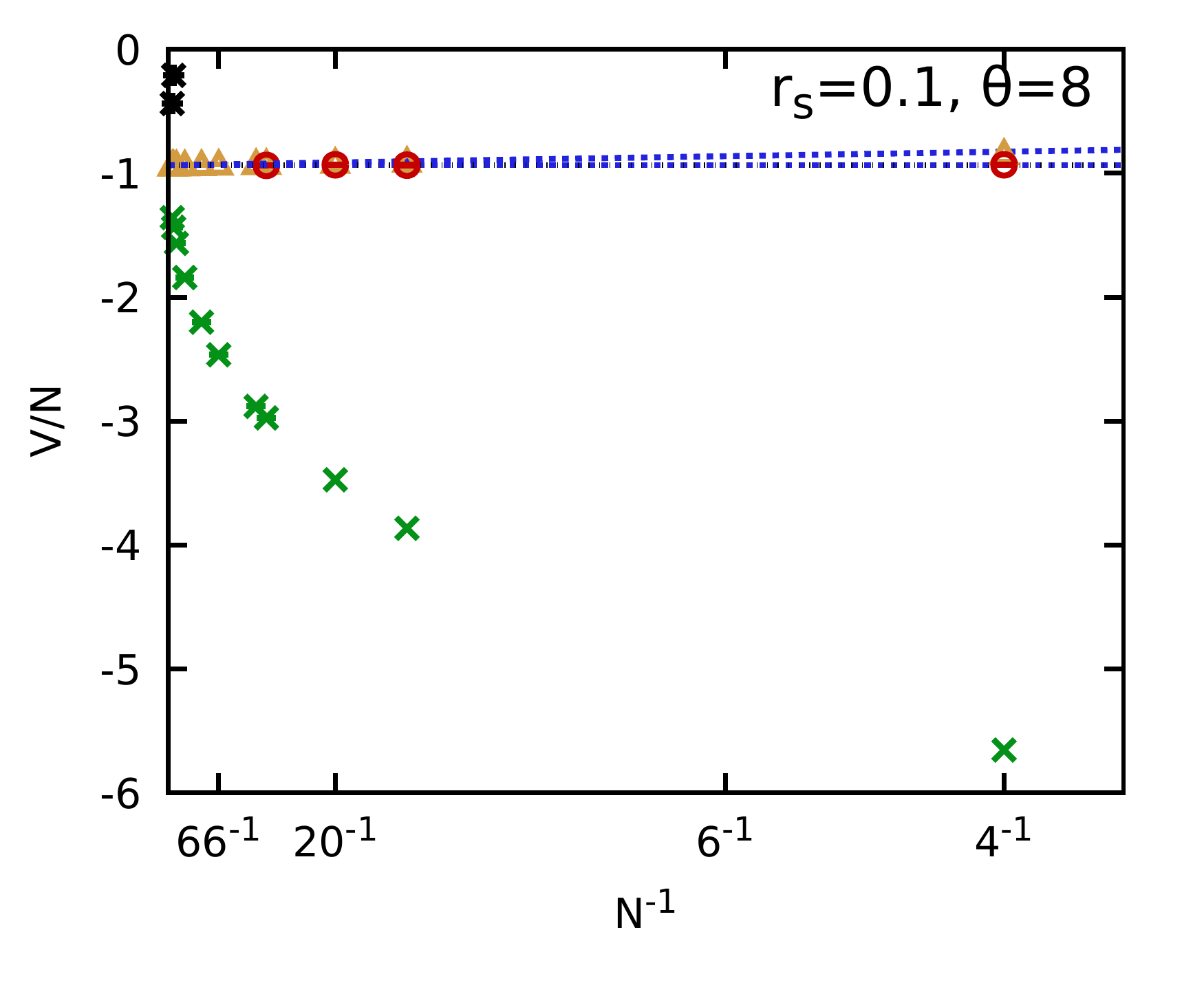}
\caption{\label{fig:v_rs0p1_theta8}
Interaction energy $v$ for $r_s=0.1$ and $\theta=8$. Top and bottom: zoomed and full view of $N$-dependence. Green crosses: raw CPIMC and PIMC data; grey stars: first-order correction $\Delta v^N_{\textnormal{d},0}$, Eq.~(\ref{eq:BCDC}); yellow triangles: $\Delta v^N_\textnormal{d}$, Eq.~(\ref{eq:FSC_v_d}); red:  $\Delta v^N_\textnormal{d}+\Delta v^N_\textnormal{i}$, Eq.~(\ref{eq:FSC_v_i}); dotted blue: linear extrapolation; 
}
\end{figure}

To discern the limits of our new FSC scheme, we next consider a substantially more extreme case, where even thousands of particles are not sufficient to resemble the TDL. In particular, we show the interaction energy per particle in Fig.~\ref{fig:v_rs0p1_theta8} for $r_s=0.1$ and $\theta=8$.
Let us first consider the bottom panel showing the entire relevant energy range of the uncorrected data, which again have been partly taken from Ref.~\cite{dornheim_prl} and computed with CPIMC, and partly obtained using standard PIMC in the context of the present work.
The raw QMC results (green crosses) naturally exhibit severe finite-size effects that exceed $500\%$ for $N=4$. Moreover, the dependence of $V^N/N$ is substantial even for $N=2000$ (the largest $N$ shown in Fig.~\ref{fig:v_rs0p1_theta8}) and does not follow any obvious functional form.
The yellow triangles have again been obtained by removing the discretization error $\Delta v^N_\textnormal{d}$ estimated using $S^\textnormal{STLS}(q)$, which reduces the dependence on $N$ by at least two orders of magnitude. Yet, there still remains a bias of several per cent for the smallest depicted values of $N$. The first-order correction $\Delta v^N_{\textnormal{d},0}$, on the other hand, is not appropriate even for $N\sim10^3$ and only becomes accurate for even larger $N$.
Finally, the red circles have been computed by subtracting from the yellow triangles the finite-size error due to the intrinsic dependence on $N$ of $S^N(q)$, $\Delta v^N_\textnormal{i}$, computed with our new scheme. Even for this most extreme case, where thousands of particles do not suffice for an appropriate description, we can predict the interaction energy in the TDL with a relative accuracy of $\sim0.2\%$ from just $N=4$ electrons.

\subsection{Increasing the coupling strength\label{sec:coupling}}

\begin{figure}\centering\includegraphics[width=0.465\textwidth]{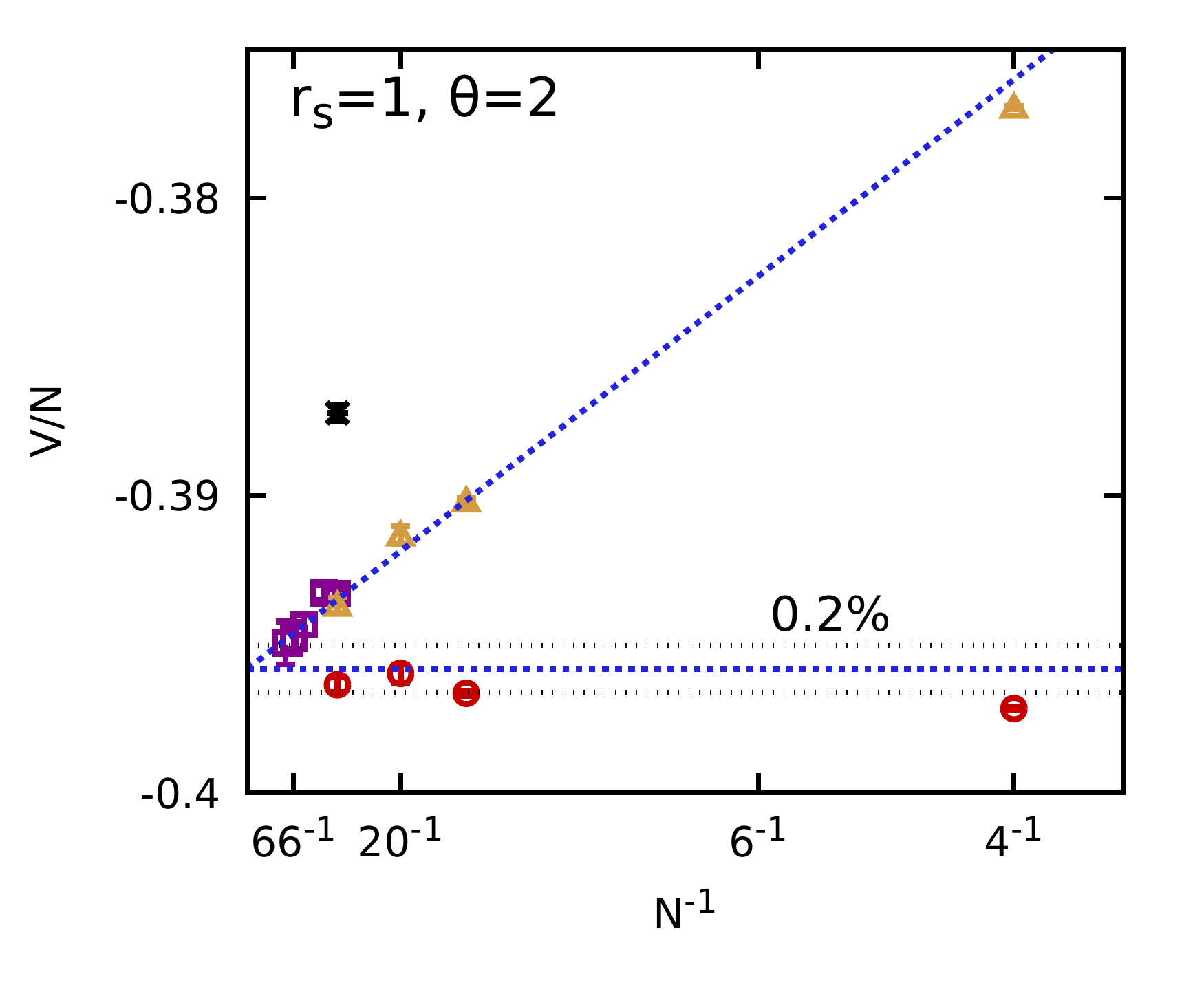}\\\vspace{-1.315cm}
\includegraphics[width=0.465\textwidth]{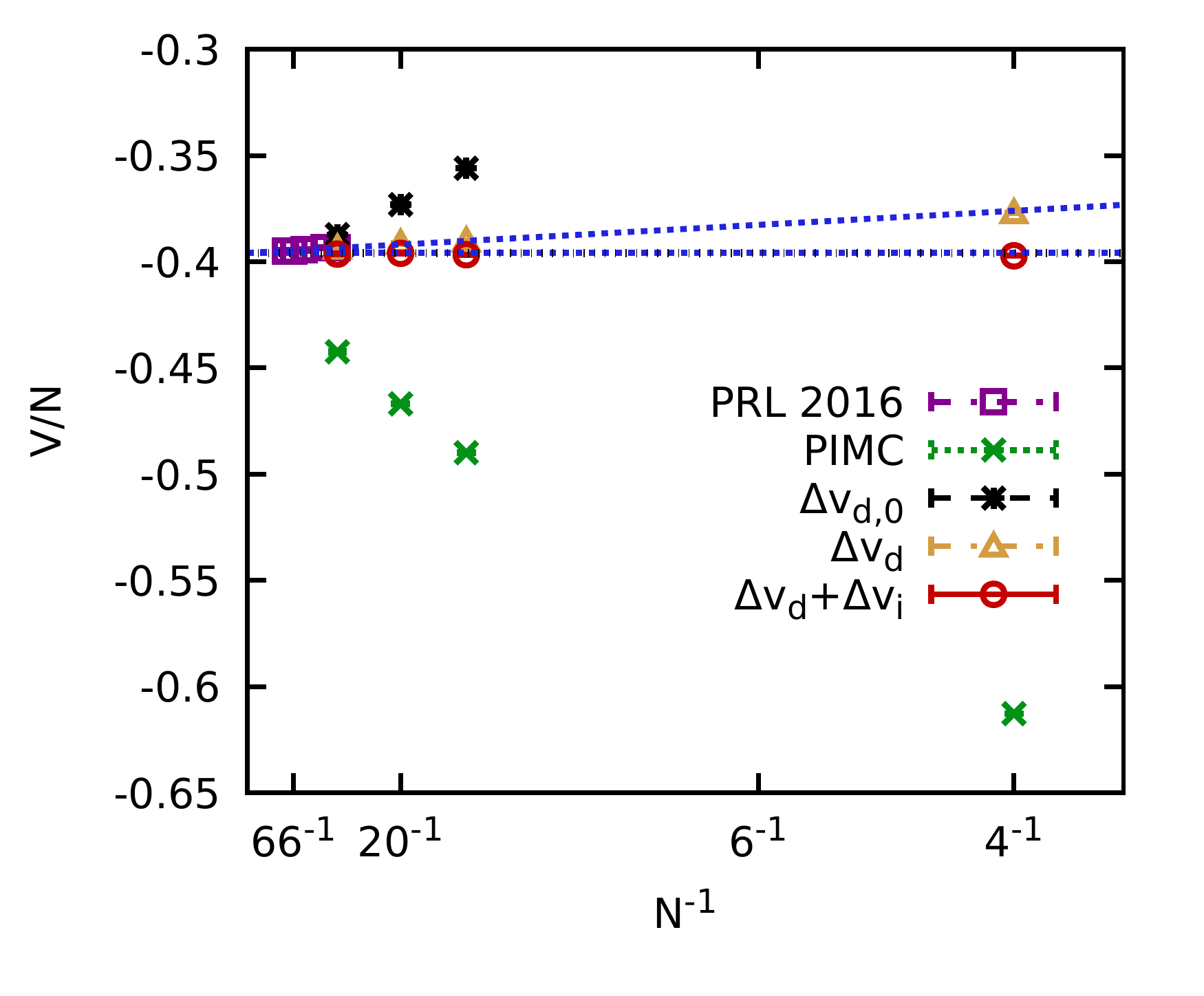}
\caption{\label{fig:v_rs1_theta2}
Interaction energy $v$ for $r_s=1$ and $\theta=2$. Top and bottom: zoomed and full view of $N$-dependence. Green crosses: raw PIMC data; grey stars: first-order correction $\Delta v^N_{\textnormal{d},0}$, Eq.~(\ref{eq:BCDC}); yellow triangles: $\Delta v^N_\textnormal{d}$, Eq.~(\ref{eq:FSC_v_d}); red:  $\Delta v^N_\textnormal{d}+\Delta v^N_\textnormal{i}$, Eq.~(\ref{eq:FSC_v_i}); dotted blue: linear extrapolation; purple squares: PB-PIMC~\cite{Dornheim_NJP_2015,dornheim_jcp} data corrected with $\Delta v^N_\textnormal{d}$ taken from Ref.~\cite{dornheim_prl}.
}
\end{figure}

To further assess the remarkable performance of our new FSC scheme demonstrated at high densities in the previous section, we will now investigate the manifestation of finite-size effects upon increasing the density parameter $r_s$. For a quantum system such as the UEG, this is equivalent to increasing the coupling strength, see e.g. Refs.~\cite{quantum_theory,Ott2018}.
A first example is shown in Fig.~\ref{fig:v_rs1_theta2}, where we plot the interaction energy per particle for $r_s=1$ and $\theta=2$. First and foremost, we find that finite-size errors are overall reduced in magnitude compared to $r_s=0.3$ shown in Fig.~\ref{fig:v_rs0p3_theta2} above, as it is expected.
In addition, the discretization error again accounts for the bulk of finite-size effects, as the yellow triangles constitute a substantial improvement over the green crosses. Further, the purple squares taken from Ref.~\cite{dornheim_prl} have been obtained using the Permutation blocking PIMC (PB-PIMC) method~\cite{Dornheim_NJP_2015,dornheim_jcp,dornheim_cpp,Dornheim_CPP_2019} and, too, were corrected using $\Delta v^N_\textnormal{d}[S^\textnormal{STLS}(q)]$. These data are in excellent agreement to the independent standard PIMC data obtained for the present work, which further corroborates the high quality of the existing description of the UEG~\cite{groth_prl,review}.
At the same time, the first-order correction $\Delta v^N_{\textnormal{d},0}$ is not sufficiently accurate to replace the full evaluation of Eq.~(\ref{eq:FSC_v_d}). Finally, the new FSC-scheme for the intrinsic error $\Delta v^N_\textnormal{i}$ results in the red circles, where the residual error does not exceed $\sim0.3\%$ even for $N=4$.

\begin{figure}\centering\includegraphics[width=0.465\textwidth]{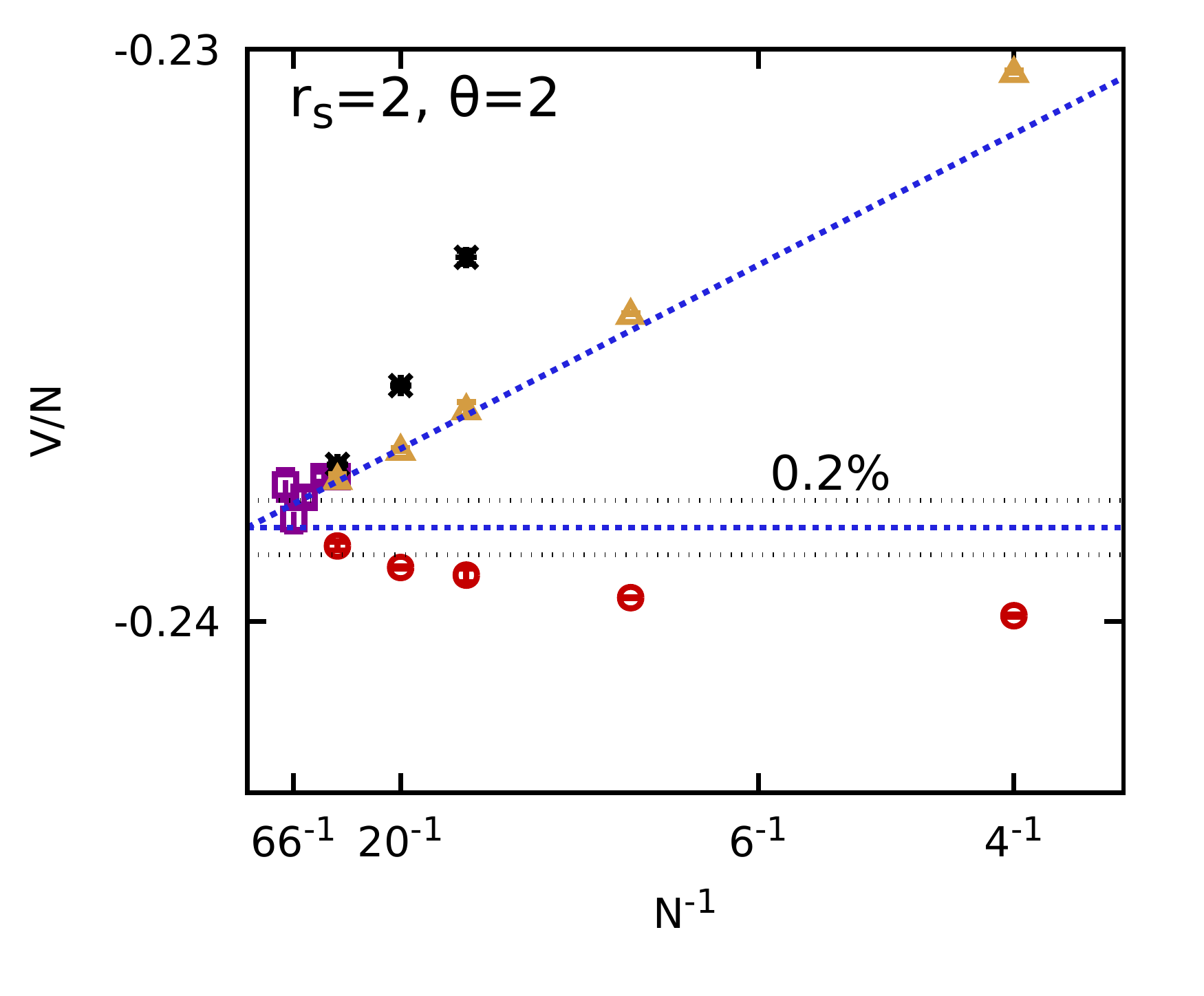}\\\vspace*{-1.35cm}
\includegraphics[width=0.465\textwidth]{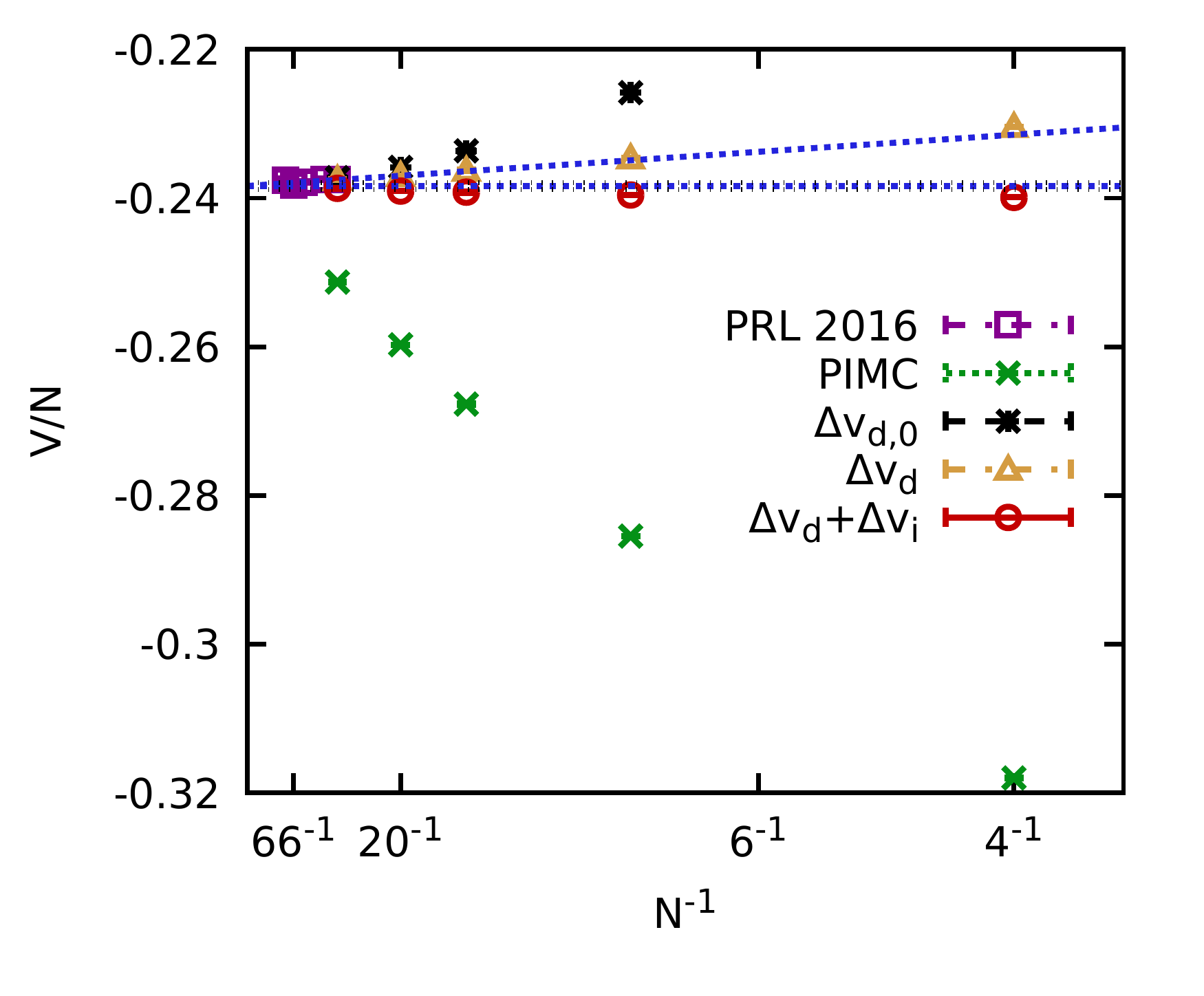}
\caption{\label{fig:v_rs2_theta2}
Interaction energy $v$ for $r_s=2$ and $\theta=2$. Top and bottom: zoomed and full view of $N$-dependence. Green crosses: raw PIMC data; grey stars: first-order correction $\Delta v^N_{\textnormal{d},0}$, Eq.~(\ref{eq:BCDC}); yellow triangles: $\Delta v^N_\textnormal{d}$, Eq.~(\ref{eq:FSC_v_d}); red circles:  $\Delta v^N_\textnormal{d}+\Delta v^N_\textnormal{i}$, Eq.~(\ref{eq:FSC_v_i}); dotted blue: linear extrapolation; purple squares: PB-PIMC~\cite{Dornheim_NJP_2015,dornheim_jcp} data corrected with $\Delta v^N_\textnormal{d}$ taken from Ref.~\cite{dornheim_prl}.
}
\end{figure}

Let us next further increase the density parameter to $r_s=2$, which is shown in Fig.~\ref{fig:v_rs2_theta2}. This is a typical density for the conduction band electrons in metals, and thus such a case is of paramount importance. In warm dense matter research, such states are created in many experiments, e.g. using aluminum~\cite{Sperling_PRL_2015,aluminum1,Takada_PRL_2002,Ramakrishna_PRB_2021}.
In this case, finite-size effects are again reduced in relative importance, and the first-order correction from Eq.~(\ref{eq:BCDC}) becomes accurate for $N\sim34$. Yet, there still remain significant residual errors in $V/N$ after the discretization error is removed, which are of the order of a few per cent for $N=4$ and decrease to $\sim0.5\%$ for $N=20$. While it is certainly feasible to carry out a controlled extrapolation of the remaining $N$-dependence of the data at these conditions, it is still worth investigating if our new scheme makes this effort superfluous.

Thus, we have added our estimation of $\Delta v^N_\textnormal{i}$ to the yellow triangles, and the results are as usual depicted by the red circles. We find that our scheme significantly reduces the dependence on $N$ for all numbers of electrons, although the actual error seems to be overestimated by Eq.~(\ref{eq:FSC_v_i}). For example, the thus fully corrected value for the interaction energy per particle for $N=4$ deviates by almost $1\%$ from the extrapolated result (horizontal dotted blue line), and these residual deviations decrease with increasing $N$.
We thus conclude that our new scheme does indeed constitute an improvement over the previous state-of-the-art of FSCs, but the finite-size error cannot be completely removed for $N=4$.

\begin{figure}\centering
\includegraphics[width=0.465\textwidth]{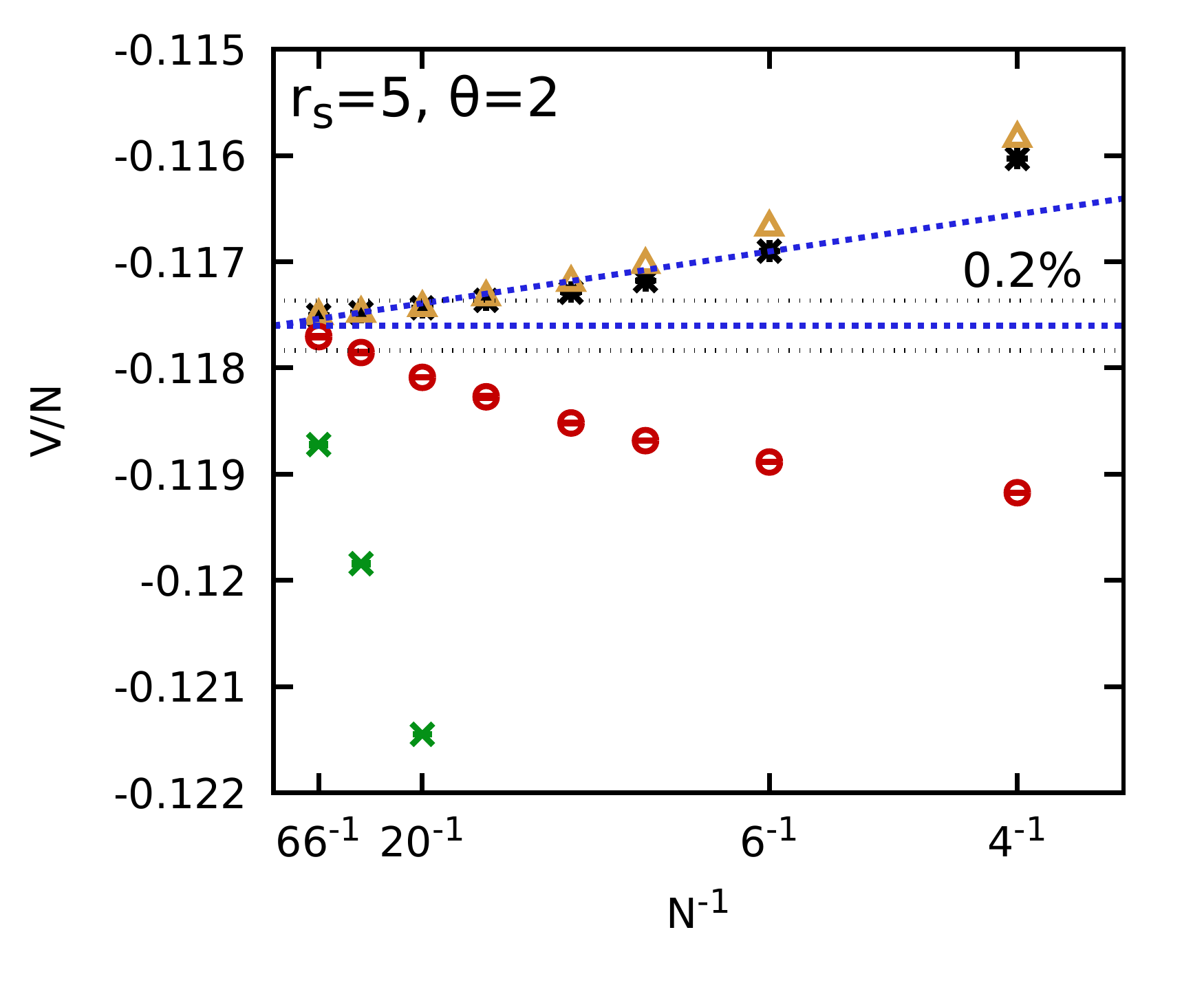}
\caption{\label{fig:v_rs5_theta2}
$N$-dependence of the interaction energy $v$ for $r_s=5$ and $\theta=2$. Green crosses: raw PIMC data; grey stars: first-order correction $\Delta v^N_{\textnormal{d},0}$, Eq.~(\ref{eq:BCDC}); yellow triangles: $\Delta v^N_\textnormal{d}$, Eq.~(\ref{eq:FSC_v_d}); red circles:  $\Delta v^N_\textnormal{d}+\Delta v^N_\textnormal{i}$, Eq.~(\ref{eq:FSC_v_i}); dotted blue: linear extrapolation; solid grey: STLS~\cite{stls_original,stls,stls2}.
}
\end{figure}

In order to understand this observed trend, it makes sense to go to even larger values of the coupling strength. To this end, we show the interaction energy per particle for $r_s=5$ and $\theta=2$ in Fig.~\ref{fig:v_rs5_theta2}. We note that such a low density serves a valuable laboratory for the impact of electronic exchange--correlation effects on material properties~\cite{low_density1,low_density2} and can be realized experimentally for example via hydrogen jets~\cite{Zastrau}.
First and foremost, we observe a further reduced relative magnitude of finite-size effects, as can be seen by the green crosses depicting the raw PIMC data for $N\geq20$. Moreover, both the full expression for $\Delta v_\textnormal{d}^N$ and the first-order term $\Delta v_{\textnormal{d},0}^N$ give fairly similar results and are all but indistinguishable for $N\gtrsim10$. Moreover, the residual error after removing the discretization error does not exceed $0.3\%$ in this case.
Finally, the red circles have been obtained by further adding to the yellow triangles the result for $\Delta v^N_\textnormal{i}$. Evidently, Eq.~(\ref{eq:FSC_v_i}) substantially overestimates true residual error and, in fact, exhibits an even somewhat more pronounced and less trivial dependence on $N$. The rest of this section will attempt to explain this unexpected break down of our expression for $\Delta v^N_\textnormal{i}$ in some detail.

\begin{figure}\centering
\includegraphics[width=0.465\textwidth]{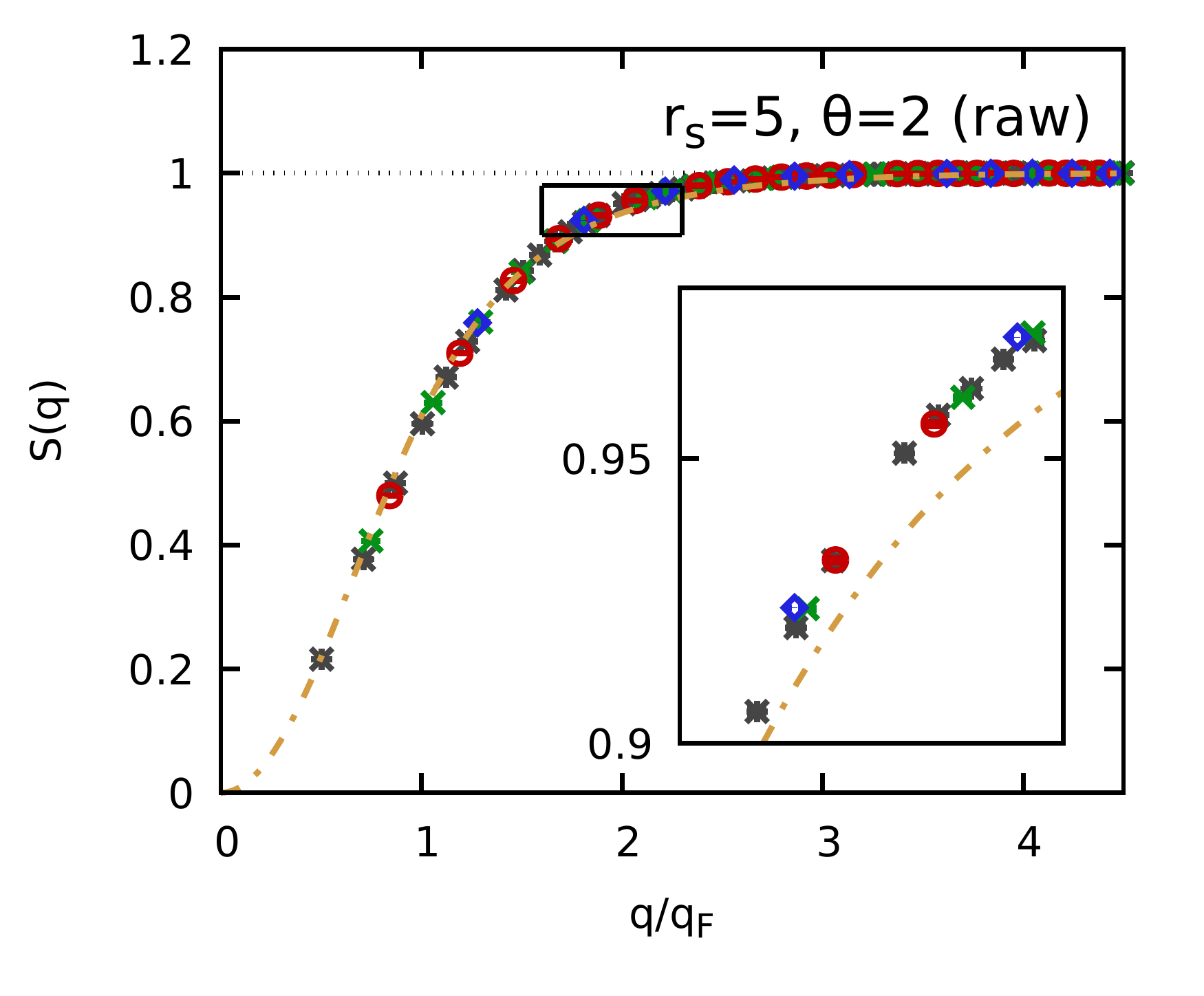}\\ \vspace*{-1.315cm}
\includegraphics[width=0.465\textwidth]{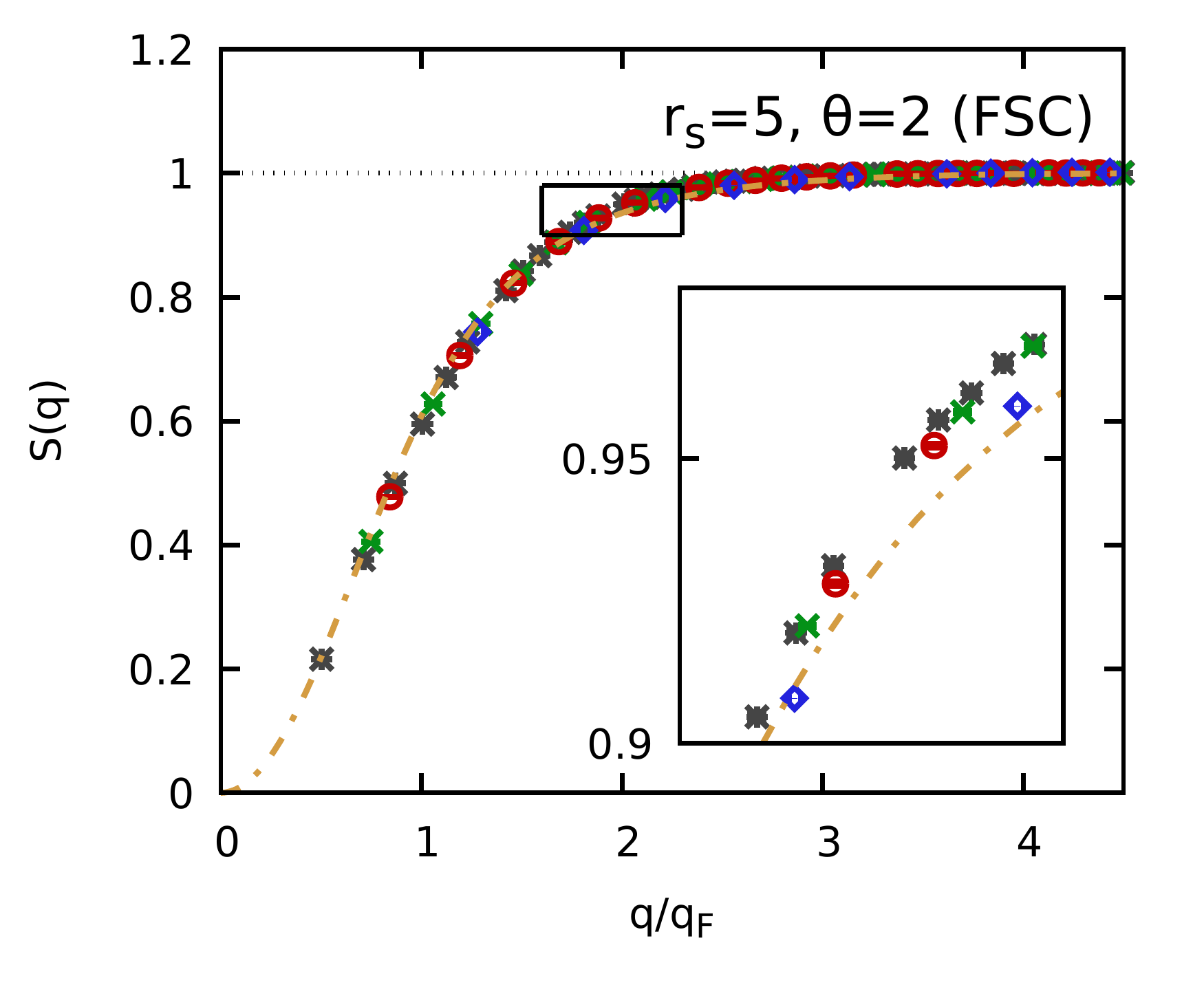}
\caption{\label{fig:SSF_rs5_theta2}
Static structure factor $S(q)$ for $r_s=5$ and $\theta=2$. The top and bottom panels show raw, uncorrected PIMC data for $S^N(q)$ and the corrected $S^N_\textnormal{FSC}(q)$ [see Eq.~(\ref{eq:FSC_S})], respectively. Blue diamonds: $N=4$; red circles: $N=14$; green crosses: $N=20$; grey stars: $N=66$. Dash-dotted yellow: STLS~\cite{stls,stls2}.
}
\end{figure} 

Let us begin this analysis by considering the central ingredient to the interaction energy, i.e., the static structure factor $S(q)$ shown in Fig.~\ref{fig:SSF_rs5_theta2}. The top panel shows standard PIMC results for the raw, uncorrected $S^N(q)$ for $N=4$ (blue diamonds), $N=14$ (red circles), $N=20$ (green crosses), and $N=66$ (grey stars). In addition, the dash-dotted yellow curve corresponds to $S^\textnormal{STLS}(q)$ and has been included as a reference. Interestingly, the only substantial intrinsic dependence of $S^N(q)$ on $N$ appears for $N=4$, whereas the other data seem to fall on a continuous curve. 

The bottom panel of the same figure shows the same information, but for the finite-size corrected version of the SSF, $S^N_\textnormal{FSC}(q)$. In this case, we find that adding our expression for $\Delta S^N(q)$ from Eq.~(\ref{eq:FSC_S}) onto $S^N(q)$ leads to a substantial increase of finite-size effects and we find a significant difference even between $N=20$ and $N=66$ that was absent in the top panel.
We have thus found, that our new scheme for the FSC for $v$ breaks down, because the involved expression for $\Delta S^N(q)$ becomes inappropriate.

\begin{figure}\centering
\includegraphics[width=0.465\textwidth]{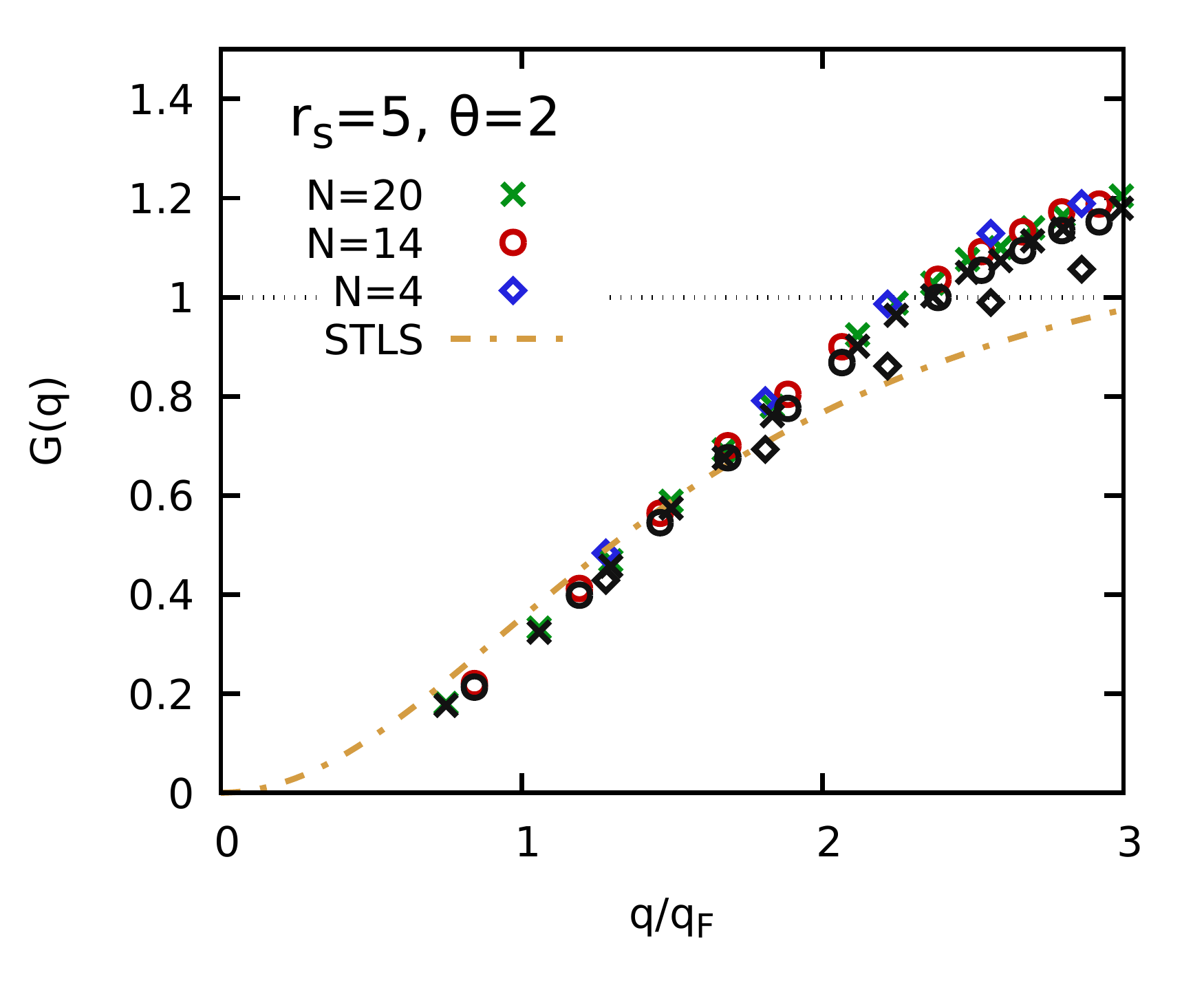}\\\vspace*{-1.33cm}
\includegraphics[width=0.465\textwidth]{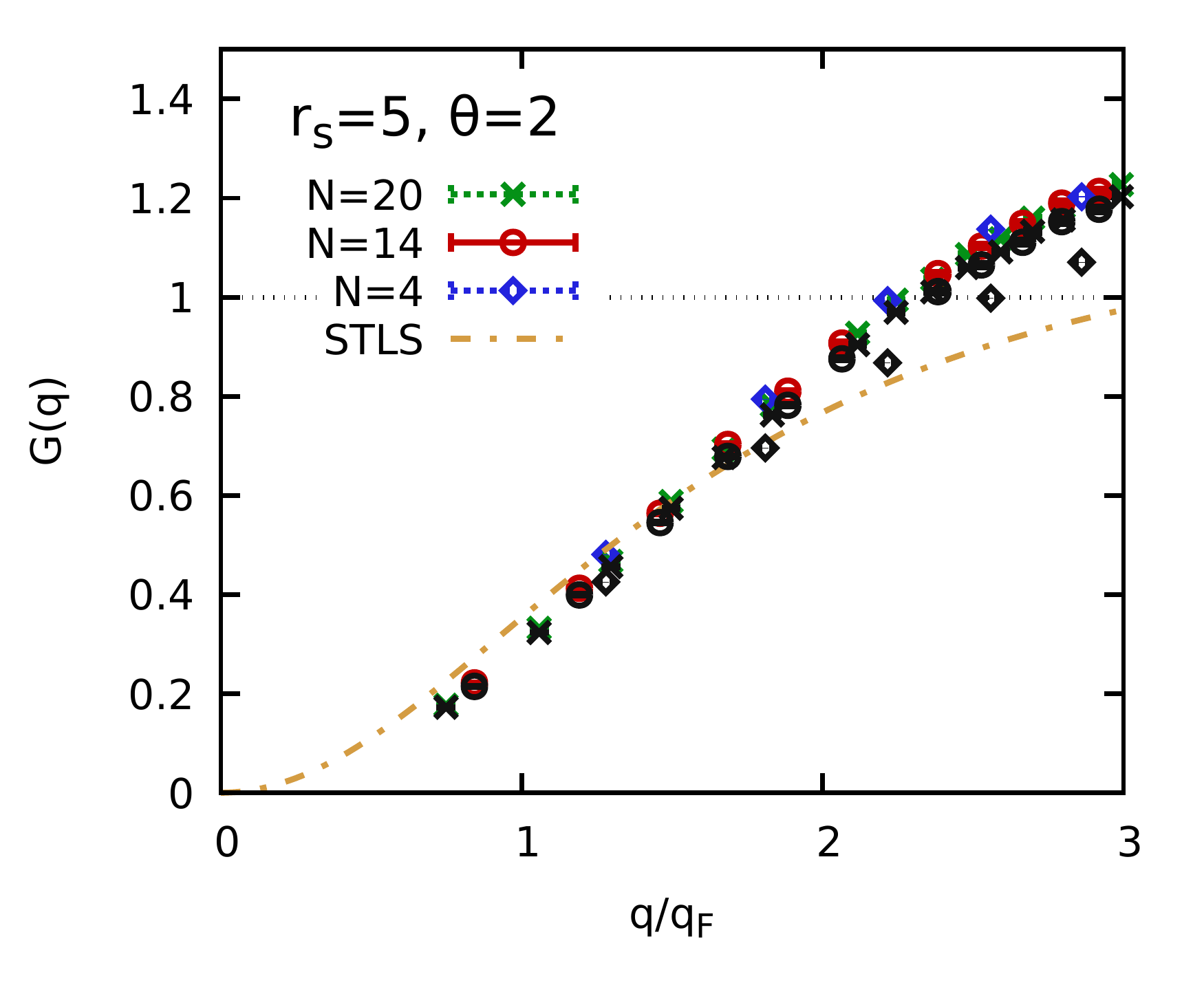}
\caption{\label{fig:G_rs5_theta2}
Top: Effective local field correction $\overline{G}^N_\textnormal{invert}(q)$ (top panel) and static limit $G^N(q,0)$ (bottom panel) for $r_s=5$ and $\theta=2$; coloured symbols: raw (uncorrected) results, see Eqs.~(\ref{eq:invert}) and (\ref{eq:G_static_N}), black: FSC-corrected, using Eq.~(\ref{eq:FSC_for_G}). Dash-dotted yellow: LFC from STLS~\cite{stls,stls2}.}
\end{figure}

Proceeding to the next deeper level of explanation, we show different local field corrections in Fig.~\ref{fig:G_rs5_theta2} for the same conditions. The top panel corresponds to the effectively static local field correction $\overline{G}_\textnormal{invert}(q)$ defined in Eq.~(\ref{eq:invert}) above, and the blue diamonds, red circles, and green crosses show the raw, uncorrected data for $N=4$, $N=14$, and $N=20$, respectively. In addition, the dash-dotted yellow line shows the LFC from STLS and has been included as a reference. Similar to our previous observation regarding $S^N(q)$ above, here, too, we only find small finite-size effects for $N=4$, and no dependence on $N$ is visible for $N=14$ and $N=20$.
The black symbols in the same panel have been obtained by adding to $\overline{G}^N_\textnormal{invert}(q)$ the finite-size correction for the static LFC $G^N(q,0)$, $\Delta G^N(q)$, given in Eq.~(\ref{eq:FSC_for_G}). Again, the FSC exacerbates the actual $N$-dependence of the uncorrected data.

This can potentially be explained by two distinctly different effects: i) the FSC for the static LFC $G^N(q,0)$ could be fundamentally different to the real, a-priori unknown FSC for $\overline{G}^N_\textnormal{invert}(q)$, or ii) our expression for $\Delta G^N(q)$ is inappropriate for both $G^N(q,0)$ and $\overline{G}^N_\textnormal{invert}(q)$.
This question is resolved by the analysis presented in the bottom panel of Fig.~\ref{fig:G_rs5_theta2}, where we show the exact static limit of the LFC for the same conditions as in the top panel. Evidently, the different data for $G(q,0)$ exhibit the same trend as $\overline{G}_\textnormal{invert}^N(q)$. In particular, the FSC $\Delta G^N(q)$ does not constitute an improvement over the uncorrected PIMC data, which means that explanation ii) holds: we do not have an appropriate theory for finite-size effects in both $G^N(q,0)$ and $\overline{G}_\textnormal{invert}^N(q)$ at these conditions.

\begin{figure}\centering
\includegraphics[width=0.465\textwidth]{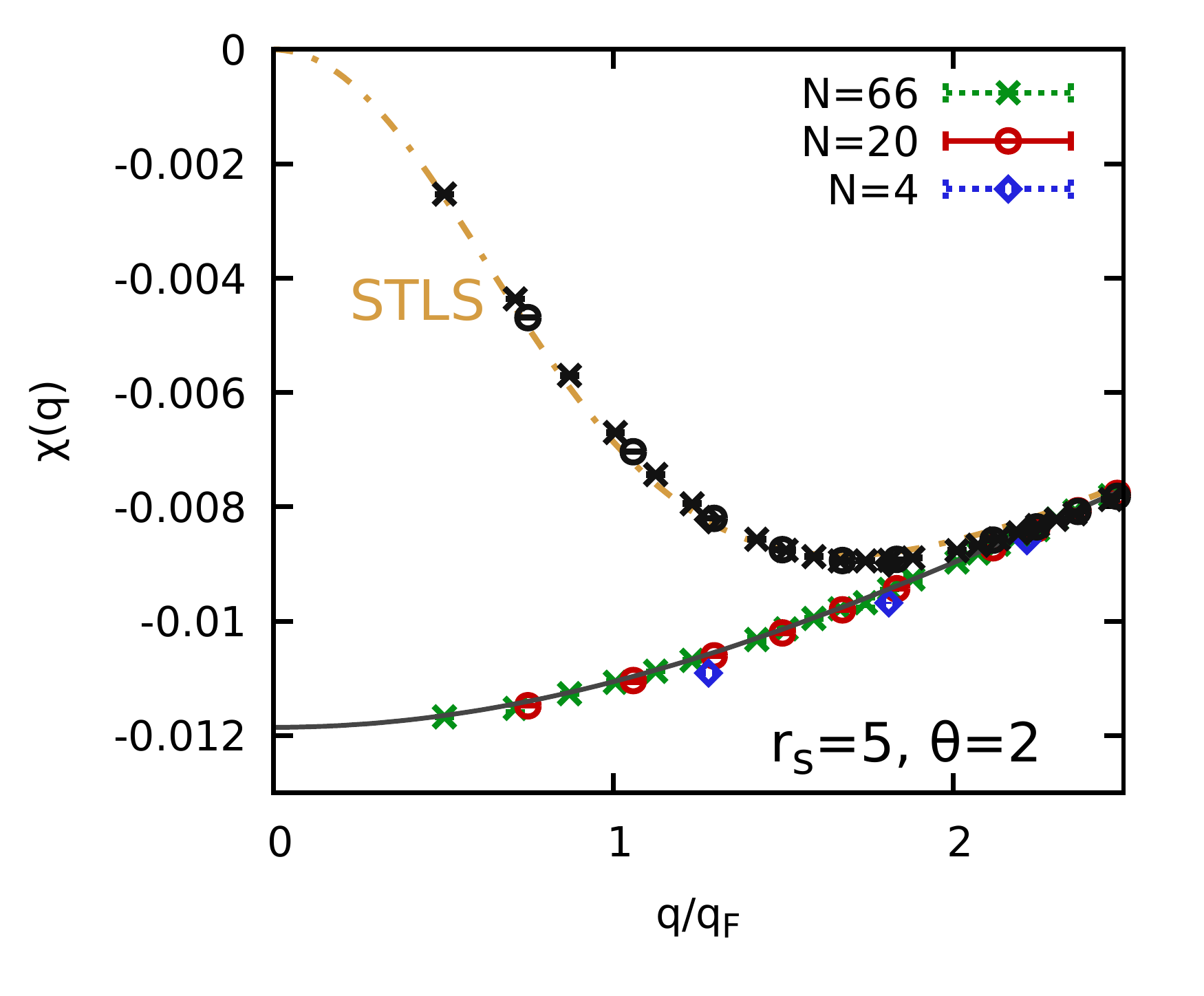}
\caption{\label{fig:Chi_rs5_theta2}
Static density response function for $r_s=5$ and $\theta=2$. Green crosses, red circles, and blue diamonds: $\chi^N_0(q)$ for $N=4$, $N=20$, and $N=66$, respectively. The corresponding black symbols show PIMC results for $\chi^N(q)$ evaluated from Eq.~(\ref{eq:static_chi}). Solid grey: $\chi^\textnormal{TDL}_0(q)$; Dash-dotted yellow: $\chi^\textnormal{STLS}(q)$~\cite{stls,stls2}.
}
\end{figure}

This can be understood more intuitively by examining Fig.~\ref{fig:Chi_rs5_theta2}, where we show data for different density response functions $\chi(q)$. More specifically, The coloured symbols depict the results for $\chi_0^N(q)$, and the solid dark grey curve depicts $\chi^\textnormal{TDL}_0(q)$. We note that there are substantial finite-size effects, in particular for $N=4$, as it is expected. In contrast, the black symbols depict the static density response function of the UEG that was computed using standard PIMC via Eq.~(\ref{eq:static_chi}), and no dependence on $N$ can be resolved with the bare eye. We have thus found that finite-size effects in the actual density response function $\chi^N(q)$ [and also in $S^N(q)$] disappear with increasing $r_s$, which does not hold for $\chi_0^N(q)$.
Yet, if it is $\chi^N(q)\approx\chi^\textnormal{TDL}(q)$, then using $\chi^\textnormal{TDL}_0(q)$ in Eq.~(\ref{eq:G_static_N}) is actually consistent and directly leads to the static local field correction in the TDL, $G^N(q)\approx G^\textnormal{TDL}(q)$, just as we have observed in Fig.~\ref{fig:G_rs5_theta2}.

The spurious effects of the finite-size correction $\Delta^N G(q)$ can then finally be interpreted in the following way: The mean-field description that is directly based on $\chi_0^N(q)$ predicts substantial finite-size effects. These, however, are absent from the actual density response function $\chi^N(q)$ which we computed via PIMC. Therefore, we can conclude that the system-size dependence is suppressed by the increased coupling strength, since the particles are considerably more localized due to the Coulomb repulsion. It is then only meaningful to define the LFC as the difference between $\chi_0^\textnormal{TDL}(q)$ and $\chi^N(q)$, as the finite-size effects in $\chi^N_0(q)$ do not manifest in the latter. If we instead would define $G^N(q,0)$ as the difference between $\chi^N(q)$ and $\chi_0^N(q)$, this LFC would have to balance the system-size dependence from $\chi^N_0(q)$, so that the actual density response function remains independent of $N$. This then explains the unreasonable behaviour of $\Delta G^N(q)$ shown in Fig.~\ref{fig:G_rs5_theta2}.

In a nutshell, our new FSC scheme implicitly assumes that the description of the system can be decomposed into a mean-field part that exhibits finite-size effects, and a short-range LFC that does not. At large values of $r_s$, this assumption breaks down, as the finite-size effects in $\chi^N_0(q)$ are suppressed by the strong Coulomb repulsion. At the same time, we note that this is not catastrophic, as the dependence of QMC data on $N$ is small for lower densities in the first place~\cite{dornheim_prl,review}.

\subsection{Going to low temperature\label{sec:temperature}}

\begin{figure}\centering\hspace*{-0.0135\textwidth}\includegraphics[width=0.48\textwidth]{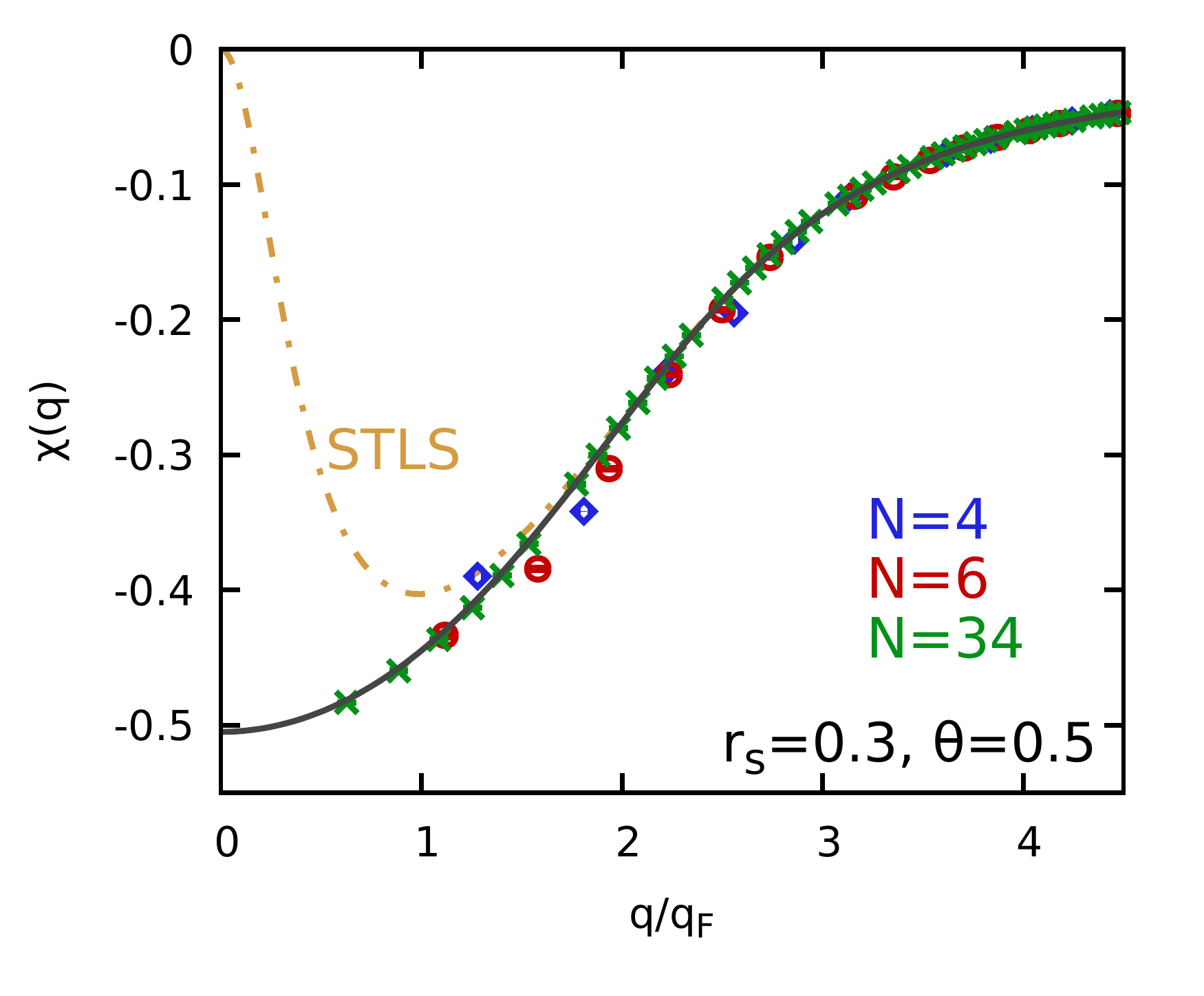}
\caption{\label{fig:Chi_rs0p3_theta0p5}
Density response functions for $\theta=0.5$ and $r_s=0.3$. The blue diamonds, red circles, and green crosses correspond to $\chi^N_0(q)$ for $N=4$, $N=6$, and $N=34$, respectively. Dark grey solid line: $\chi^\textnormal{TDL}_0(q)$; dash-dotted yellow: $\chi^\textnormal{STLS}(q)$, see Refs.~\cite{stls,stls2}.
}
\end{figure}

The final part of the investigation in this work is devoted to the performance of our new FSC scheme for low temperatures. This is a particularly interesting regime, as accurate QMC data for the UEG are sparse at these conditions due to the fermion sign problem~\cite{dornheim_sign_problem,lee2020phaseless,Yilmaz_JCP_2020,dornheim_POP}. In addition, the UEG has attracted renewed interest at high densities also in the ground-state~\cite{Shepherd_UEG_2012,Shepherd_UEG_PRB_2012}, where the full configuration interaction QMC (FCIQMC) method~\cite{Booth_JCP_2009} is capable to give accurate results for finite $N$.

One of the main sources of finite-size errors at low temperature is given by momentum-shell effects. For this reason, one typically selects a system-size that is commensurate to the grid in momentum space, e.g. for an unpolarized UEG $N=14,38,54,66,\dots$. An additional strategy is to employ periodic boundary conditions with a finite twist-angle~\cite{Lin_Zong_Ceperley_PRE_2001,Zong_Lin_Ceperley_PRE_2002,Sorella_FSC_PRB_2015,Azadi_Foulkes_PRB_2019,Shepherd_JCP_2019} and perform a subsequent \emph{twist-averaging} which has been shown to greatly alleviate these effects.
At the same time, we note that these momentum shell effects are already fully present in $\chi^N_0(q)$, which makes our new FSC promising in this regime. This is illustrated in Fig.~\ref{fig:Chi_rs0p3_theta0p5}, where we show the density response function for $r_s=0.3$ and $\theta=0.5$. Here, the blue diamonds, red circles, and green crosses show $\chi_0^N(q)$ for $N=4$, $N=6$, and $N=34$, respectively, and the solid dark grey curve corresponds to $\chi^\textnormal{TDL}_0(q)$. In addition, the dash-dotted curve shows $\chi^\textnormal{STLS}(q)$ and has been included as a reference.
Firstly, we note that $\theta=0.5$ only constitutes the boundary of the low-temperature regime, and thermal excitations still play an important role~\cite{review}. Still, incipient momentum shell effects clearly manifest for $\chi_0^N(q)$ in Fig.~\ref{fig:Chi_rs0p3_theta0p5}, in particular for $N=4$. Similar results for lower temperatures have been presented by Groth~\textit{et al.}~\cite{groth_jcp}.

\begin{figure}\centering\hspace*{-0.0135\textwidth}\includegraphics[width=0.48\textwidth]{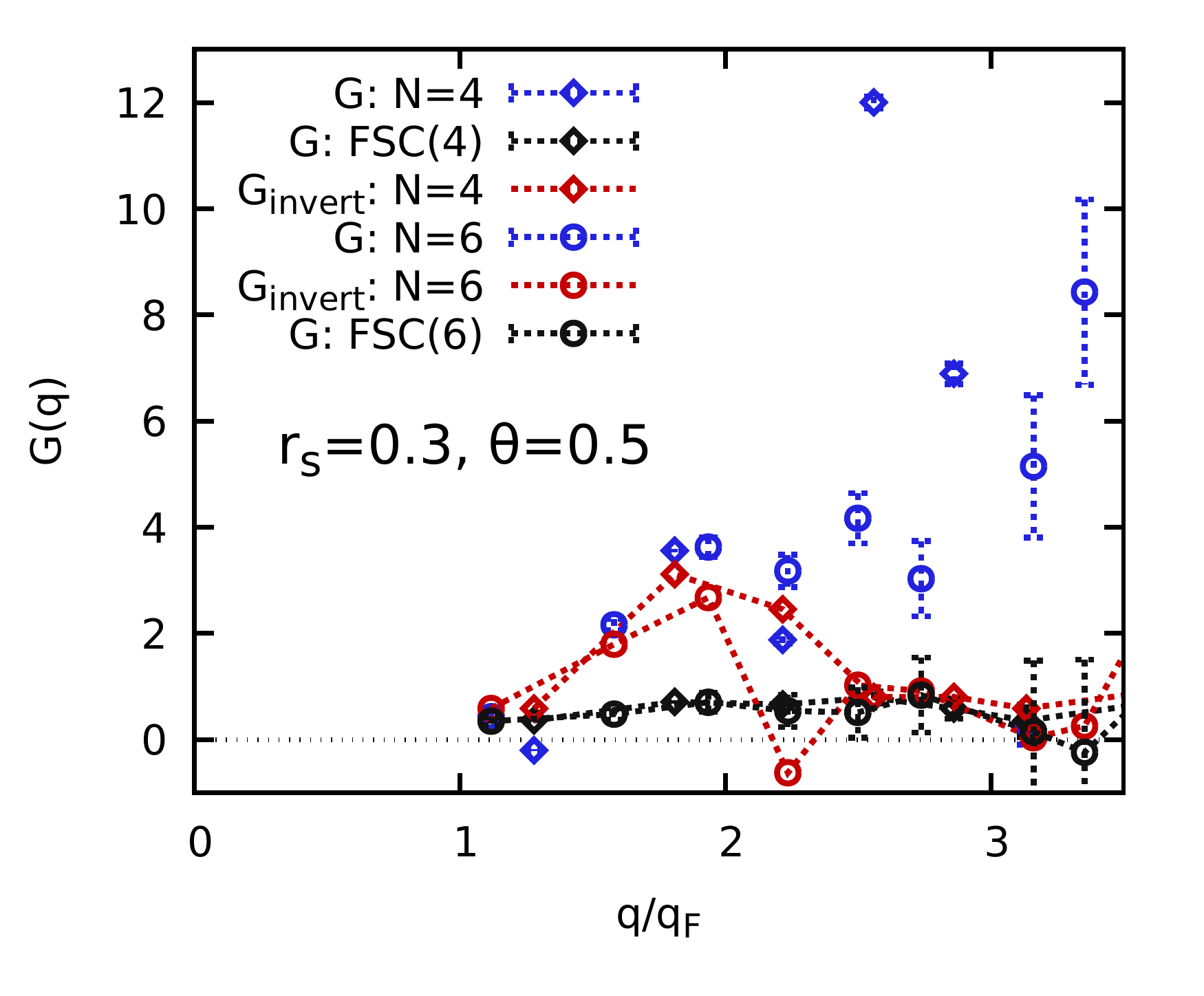}\\\vspace{-1.4cm}\hspace*{-0.025\textwidth}\includegraphics[width=0.493\textwidth]{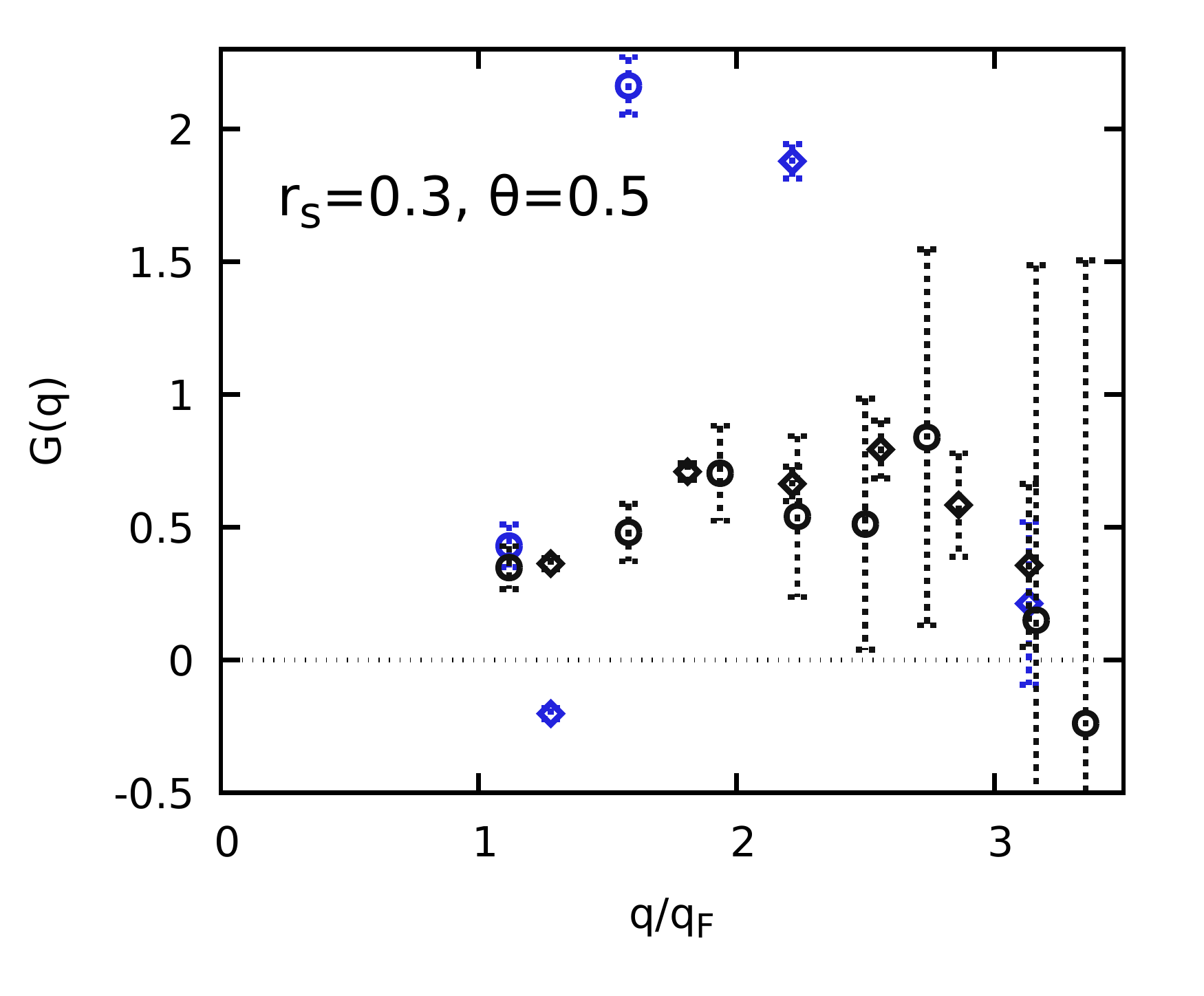}
\caption{\label{fig:LFC_rs0p3_theta0p5}
Different local field corrections for $\theta=0.5$ and $r_s=0.3$. The diamonds and circles correspond to $N=4$ and $N=6$ electrons. Blue: raw, uncorrected static LFC $G^N(q,0)$, see Eq.~(\ref{eq:G_static_N}); black: $G^\textnormal{FSC}_N(q)$ using Eq.~(\ref{eq:FSC_for_G}); red: raw, uncorrected results for $\overline{G}^N_\textnormal{invert}(q)$, see Eq.~(\ref{eq:invert}). Bottom panel: magnified view around the corrected values $G^\textnormal{FSC}_N(q)$ from the top panel.
}
\end{figure}

The next step towards a finite-size correction for $S^N(q)$ and $V^N/N$ is given by analysis of different local field corrections presented in Fig.~\ref{fig:LFC_rs0p3_theta0p5} for the same conditions. Let us first consider the top panel where we show results for $N=4$ (diamonds) and $N=6$ (circles). The red data show $\overline{G}_\textnormal{invert}^N(q)$, which constitutes the basis of our FSC scheme for $\Delta S^N(q)$ and, in turn, $\Delta v^N_\textnormal{i}$. Apparently, the data sets for the two different particle numbers do not follow a smooth progression, and a finite-size correction is needed; similarly, $S^N(q)$ exhibits finite-size effects, see the discussion of Fig.~\ref{fig:SSF_rs0p3_theta0p5} below.
The blue data show results for the static limit of the LFC $G^N(q,0)$ and have been obtained from PIMC simulations via Eq.~(\ref{eq:static_chi}) and Eq.~(\ref{eq:G_static_N}). On the one hand, these data, too, do exhibit substantial finite-size effects, which further underlines the need for an FSC. On the other hand, the system-size dependence of $G^N(q)$ does not resemble the observed behaviour of $\overline{G}^N_\textnormal{invert}(q)$, which makes the applicability of our FSC scheme questionable.

Remarkably, adding the FSC $\Delta G^\textnormal{FSC}(q)$ to the blue points gives the black curves, for which no dependence on $N$ can be resolved within the given level of statistical uncertainty. This can be seen particularly well in the bottom panel where we show a magnified view around $G^\textnormal{FSC}_N(q)$. Thus, QMC simulations of $N=4$ and $N=6$ electrons appear to give us access to $G(q,0)$ in the TDL, but not $\overline{G}^\textnormal{TDL}_\textnormal{invert}(q)$ as no theory for $\Delta \overline{G}^N_\textnormal{invert}(q)$ is at hand.

\begin{figure}\centering\hspace*{-0.0135\textwidth}\includegraphics[width=0.48\textwidth]{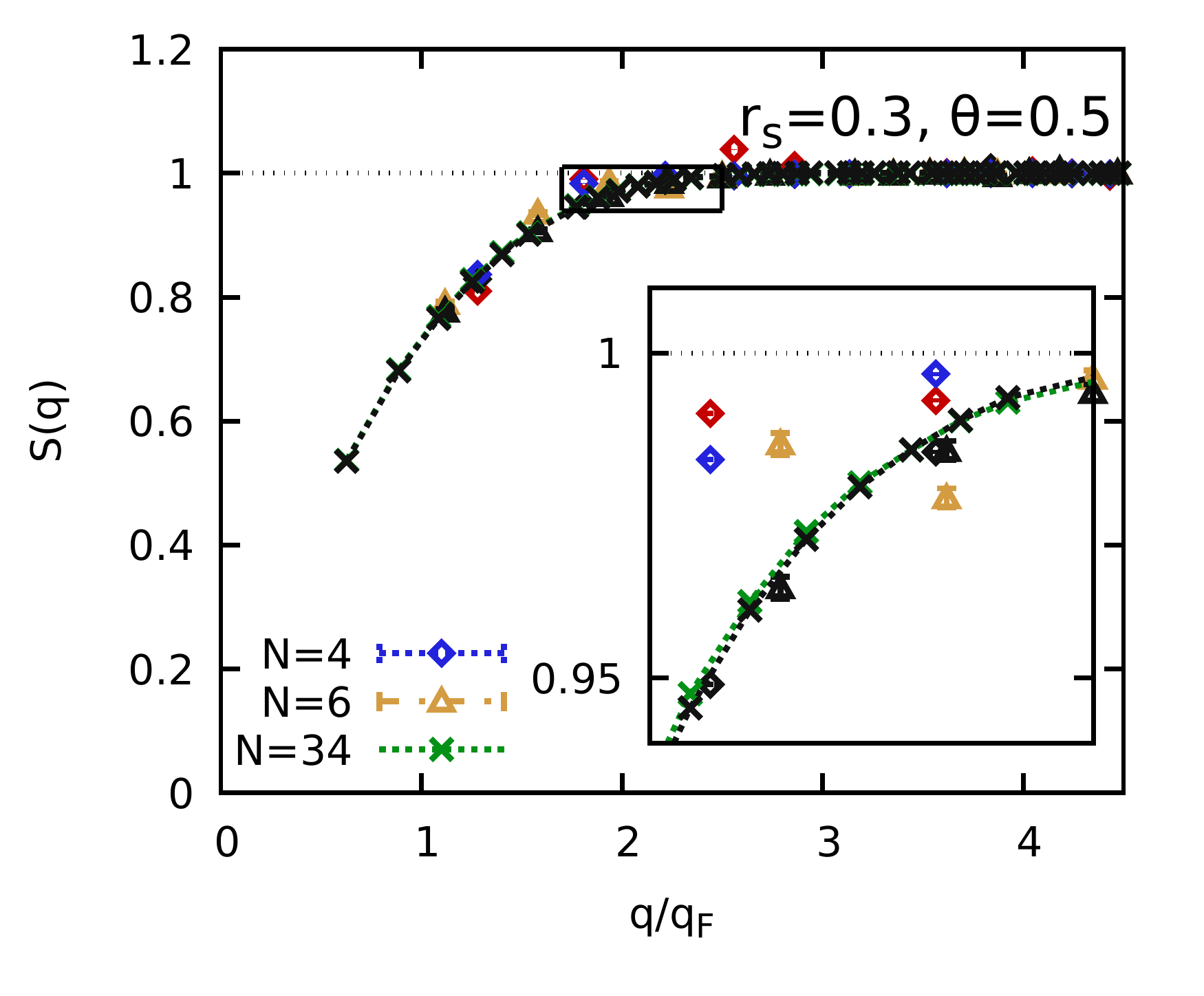}
\caption{\label{fig:SSF_rs0p3_theta0p5}
SSF for $\theta=0.5$ and $r_s=0.3$. Green crosses: raw CPIMC data for $N=34$ taken from Ref.~\cite{dornheim_prl}; blue diamonds: raw PIMC data for $N=4$; yellow triangles: raw PIMC data for $N=6$; the corresponding black symbols have been finite-size corrected, for $N=34$ in the regular way [Eq.~(\ref{eq:FSC_S})], for $N=4,6$ with Eq.~(\ref{eq:FSC_S_stat}); red diamonds: static approximation using $G^N(q,0)$ for $N=4$.
}
\end{figure}

Let us next ponder the consequences of these findings for the FSC of the static structure factor $S(q)$, which we show in Fig.~\ref{fig:SSF_rs0p3_theta0p5} for the same conditions. Here the blue diamonds, yellow triangles and green crosses show the raw, uncorrected QMC data for $S^N(q)$.
We note that the observed dependence on $N$ is qualitatively similar to the finite-size effects observed in $\chi_0^N(q)$ in Fig.~\ref{fig:Chi_rs0p3_theta0p5} above. In addition, the red diamonds have been obtained by using $G^N(q,0)$ to compute the static structure factor within the \emph{static approximation} [Eq.~(\ref{eq:static_approximation})]. Evidently, these data do substantially disagree with the actual QMC data for $S^N(q)$, which appears to indicate that the \emph{static approximation} is not applicable at these conditions. This is not surprising and directly explains the difference between $\overline{G}^N_\textnormal{invert}(q)$ and $G^N(q)$ in Fig.~\ref{fig:LFC_rs0p3_theta0p5}.

Yet, the full picture is even more subtle and can be deduced from the black diamonds in Fig.~\ref{fig:SSF_rs0p3_theta0p5}, which show the static structure factor $S^\textnormal{FSC}_\textnormal{stat}(q)$ obtained via the \emph{static approximation} using as input the finite-size corrected LFC $G^\textnormal{FSC}_N(q)$ for $N=4$. Remarkably, these data do not exhibit the unsmooth behaviour of $S^N(q)$ associated with the momentum shell effects, but for a smooth progression with $q$. Moreover, these data are much closer to the green crosses depicting the QMC data for $N=34$, and the black crosses, that have been obtained by applying the usual FSC $\Delta S^N(q)$ [Eq.~(\ref{eq:FSC_S})] to the latter.

These findings indicate that the \emph{static approximation} does not break down at these conditions after all, since the black diamonds appear to be accurate. Instead, the large discrepancy between the blue and red diamonds follows from an inconsistent mixing of different response functions, which did not matter for $\theta=2$, but has a large impact in the low temperature regime. More specifically, a consistent \emph{static approximation} for a finite number of electrons $N$ should have the form
\begin{eqnarray}\label{eq:static_approximation_N}
\chi^N_\textnormal{stat}(q,\omega) = \frac{\chi^N_0(q,\omega)}{1-\frac{4\pi}{q^2}\left[1-G^N(q)\right]\chi^N_0(q,\omega)}\ ,
\end{eqnarray}
where the dynamic density response function of the ideal Fermi gas is computed for the same system size. Thus, combining $G^\textnormal{FSC}_N(q)$ with $\chi^\textnormal{TDL}_0(q,\omega)$ gives reasonable results for $S(q)$ (black diamonds), whereas mixing $G^N(q)$ with $\chi^\textnormal{TDL}_0(q,\omega)$ does neither fit to the TDL, nor the QMC data for $S^N(q)$. Following this line of thinking, our definition of $\overline{G}_\textnormal{invert}^N(q)$ from Eq.~(\ref{eq:invert}) above, too, is inconsistent, as it must reproduce $S^N(q)$ using the inappropriate TDL result $\chi^\textnormal{TDL}_0(q,\omega)$. 
Instead, one should define a modified effective LFC, where the functional $S[\overline{G}(q)](q)$ is evaluated using the \emph{static approximation} from Eq.~(\ref{eq:static_approximation_N}).
The implementation of this idea, however, is beyond the scope of the present work and constitutes a project for future research.

The final question to be discussed here is whether it is still possible to compute a reasonable estimation for the intrinsic contribution to the finite-size error of the interaction energy $\Delta v^N_\textnormal{i}$ without detailed knowledge of $\chi^N_0(q,\omega)$.
This is indeed possible, as we do have access to finite-size corrected data for $S(q)$ in the form of the black diamonds in Fig.~\ref{fig:SSF_rs0p3_theta0p5} that have been obtained from the \emph{static approximation} using as input $G^\textnormal{FSC}_N(q)\approx G^\textnormal{TDL}(q)$.
From these, we can deduce an approximate expression for the finite-size effects in $S^N(q)$ as 
\begin{eqnarray}\label{eq:FSC_S_stat}
\Delta S^N_\textnormal{stat}(q) = S^\textnormal{FSC}_\textnormal{stat}(q) - S^N(q)\ .
\end{eqnarray}
In particular, we stress that Eq.~(\ref{eq:FSC_S_stat}) cannot be exact, as $S^\textnormal{FSC}_\textnormal{stat}(q)$ entails the (small) systematic error from the \emph{static approximation} [i.e., neglecting the frequency-dependence of the full LFC $G(q,\omega)$], which is absent from $S^N(q)$. Still, we stress that this systematic bias is small, especially at the high densities that are of interest to the present work~\cite{dornheim_dynamic,Hamann_PRB_2020}, and that Eq.~(\ref{eq:FSC_S_stat}) is only used to estimate the small correction to the interaction energy per particle, and not $V^N/N$ itself. The corresponding expression for the intrinsic contribution to $\Delta v^N$ is then given by
\begin{eqnarray}\label{eq:FSC_v_i_stat}
\Delta v_\textnormal{i}^N[\textnormal{stat}] = \frac{1}{2V}\sum_{\mathbf{G}\neq\mathbf{0}}\left[S_\textnormal{stat}^\textnormal{FSC}(\mathbf{G}) - S^N(\mathbf{G}) \right] \frac{4\pi}{G^2}\ .
\end{eqnarray}

\begin{figure}\centering\hspace*{-0.033\textwidth}\includegraphics[width=0.505\textwidth]{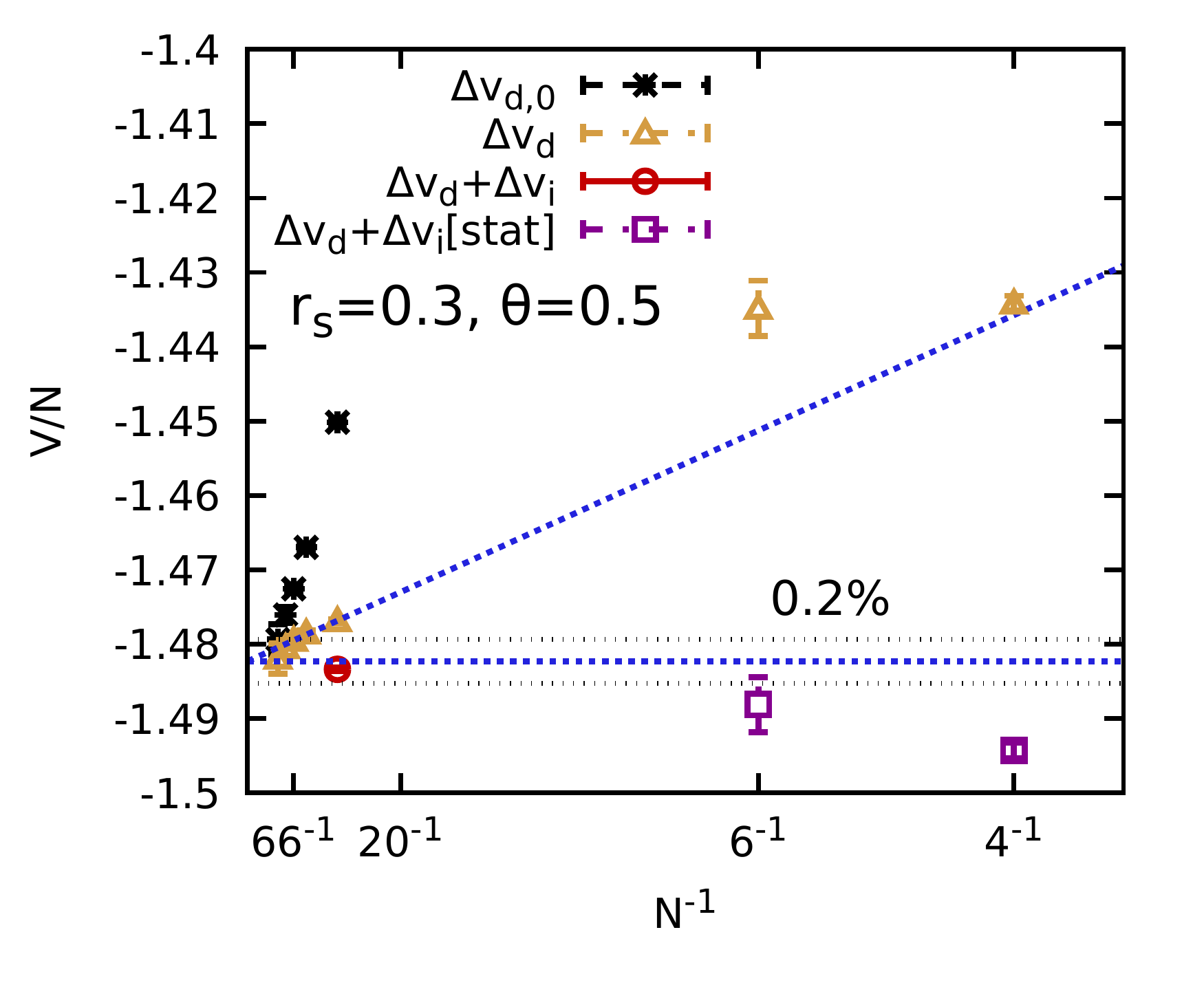}
\caption{\label{fig:v_rs0p3_theta0p5}
Interaction energy $v$ for $r_s=0.3$ and $\theta=0.5$.  Green crosses: raw CPIMC and PIMC data; grey stars: first-order correction $\Delta v^N_{\textnormal{d},0}$, Eq.~(\ref{eq:BCDC}); yellow triangles: $\Delta v^N_\textnormal{d}$, Eq.~(\ref{eq:FSC_v_d}); red circles:  $\Delta v^N_\textnormal{d}+\Delta v^N_\textnormal{i}$, Eq.~(\ref{eq:FSC_v_i}); dotted blue: linear extrapolation; purple squares: adapted FSC $\Delta v^N_\textnormal{d}+\Delta v^N_\textnormal{i}[\textnormal{stat}]$, see Eq.~(\ref{eq:FSC_v_i_stat}).
}
\end{figure}

Let us conclude our investigation of finite-size effects in electronic structure simulations at extreme conditions with an analysis of the interaction energy shown in Fig.~\ref{fig:v_rs0p3_theta0p5}. In particular, we only show a magnified segment around the finite-size corrected data sets, as the overall trend is quite similar to the case of $\theta=2$ shown in Fig.~\ref{fig:v_rs0p3_theta2} above. Further, the zero-order expression $\Delta v_{\textnormal{d},0}^N$ from Eq.~(\ref{eq:BCDC}) only becomes appropriate for large $N$, and the full estimation of $\Delta v_\textnormal{d}^N$ is needed. Still, there remains a significant residual error that only vanishes within the given Monte Carlo error bars for $N\gtrsim66$. Using Eq.~(\ref{eq:FSC_v_i_stat}) to estimate the intrinsic error in $v$ gives the purple squares, where the residual error is reduced by nearly an order of magnitude.
We thus conclude that accurate knowledge of the static response functions $\chi^N(q)$ and $\chi_0^N(q)$ can be used in lieu of the full frequency-dependence of $\chi^N_0(q,\omega)$ to compute a reasonable estimate of $\Delta v_\textnormal{i}^N$ within the \emph{static approximation} at low temperatures.

\section{Summary and Outlook\label{sec:summary_and_outlook}}

\subsection{Summary\label{sec:summary}}

In summary, we have presented a FSC scheme for the static structure factor $S(q)$ and, in this way, also the interaction energy per particle $v$. More specifically, we have employed the density response formalism and identified an effectively frequency-averaged static local field correction $\overline{G}_\textnormal{invert}(q)$ as a short-ranged exchange--correlation property that can be extracted from a QMC simulation of only a few particles.

As a practical application, we have investigated the UEG at extreme conditions~\cite{review} and demonstrated that often as few as $N=4$ electrons are sufficient to obtain the interaction energy with a relative accuracy of $\sim0.2\%$. We stress that our approach is completely nonempirical, and does not require any external input or simulation data for different $N$.

In addition, we have analyzed the applicability of our scheme upon increasing the density parameter $r_s$ (and thus the coupling strength), and found that it eventually breaks down when correlation effects dominate. Finally, we have investigated the UEG at low temperature, where momentum shell effects constitute an additional source of finite-size errors. Still, these errors, too, are present in the density response function of the noninteracting system $\chi^N_0(q)$, and we have outlined two different strategies to mitigate the $N$-dependence in this case: i) knowledge of the static interacting response function $\chi^N(q)$ allows to obtain an accurate (though not exact) FSC for both $S(q)$ and $v$ which reduces the residual error in $v$ after the elimination of the discretization error by an additional order of magnitude, and ii) using $\chi^N_0(q,\omega)$ would allow to construct a consistent, $N$-dependent implementation of the \emph{static approximation}, which has the potential to fully remove momentum shell effects and, in this way, remove the need for an additional twist-averaging procedure.

\subsection{Outlook\label{sec:outlook}}

We believe that our findings open new avenues for additional topics of research to be pursued in future investigations. First and foremost, we call to mind the direct relation between $S(q)$ and the pair distribution function $g(r)$, which can be exploited to derive a FSC for the latter. In particular, this could be vital for the determination of $g(0)$, which has most recently been investigated by Hunger \textit{et al.}~\cite{Hunger_PRE_2021} and is important for different applications, e.g.~Refs.~\cite{dornheim_PRL_ESA_2020,Dornheim_PRB_ESA_2021}. In addition, the idea behind $S(q)$ can be straightforwardly extended to the imaginary-time density--density correlation function $F(q,\tau)$ defined in Eq.~(\ref{eq:F}) above. This function is of high value, as it allows for the computation of the dynamic structure factor $S(q,\omega)$ and related properties~\cite{dornheim_dynamic,dynamic_folgepaper,Hamann_PRB_2020}, which, in turn, are accessible in experiments, e.g. using the X-ray Thomson scattering technique~\cite{siegfried_review,kraus_xrts}. Furthermore, we mention the well-known thermodynamic relations between different energies, e.g. Eq.~(\ref{eq:adiabatic}), which potentially allow to generalize our correction for $v$ to other quantities such as the kinetic energy.
Finally, we point out that our method is particularly successful for high temperatures and densities, which indicates that it may apply to classical molecular dynamics simulations~\cite{Mithen_PRE_2012,Hanno_PRR_2020,Ott_PRL_dispersion_2012} as well.

\section*{Acknowledgments}
We are grateful to S.~Groth for sending us various CPIMC data sets for $S^N(q)$ and $\chi^N_0(q)$.
This work was partly funded by the Center for Advanced Systems Understanding (CASUS) which is financed by Germany's Federal Ministry of Education and Research (BMBF) and by the Saxon Ministry for Science, Culture and Tourism (SMWK) with tax funds on the basis of the budget approved by the Saxon State Parliament.
We gratefully acknowledge CPU-time at the Norddeutscher Verbund f\"ur Hoch- und H\"ochstleistungsrechnen (HLRN) under grant shp00026 and on a Bull Cluster at the Center for Information Services and High Performace Computing (ZIH) at Technische Universit\"at Dresden.

\bibliography{bibliography}
\end{document}